\documentclass[preprint,12pt]{elsarticle}

\usepackage{amsmath}
\usepackage[linesnumbered,ruled,vlined]{algorithm2e}
\usepackage{pifont} 

\newcommand{\cmark}{\ding{51}} 
\newcommand{\xmark}{\ding{55}} 
\usepackage{longtable}

\SetCommentSty{mycommfont}

\SetKwInput{KwInput}{Input}                
\SetKwInput{KwOutput}{Output}  

\usepackage{tikz}

\usepackage{hyperref}
\hypersetup{
    colorlinks=true,
    linkcolor=blue,
    filecolor=magenta,      
    urlcolor=cyan,
    pdftitle={Overleaf Example},
    pdfpagemode=FullScreen,
    }

\usepackage{amssymb}
\usepackage{multirow}
\usepackage{threeparttable}
\usepackage{amssymb}
\usepackage{xtab}
\usepackage{xcolor}
\usepackage{threeparttablex}
\usepackage{pifont}
\setcitestyle{citesep={,}}
\usepackage{hyperref}
\usepackage{url}
\makeatletter
\g@addto@macro{\UrlBreaks}{\UrlOrds}
\makeatother

\setcitestyle{square}
\usepackage{algorithmic}
\usepackage{graphicx}
\usepackage{subcaption}
\usepackage{algorithm2e}
\usepackage{longtable}
\SetKwComment{Comment}{/* }{ */}
\RestyleAlgo{ruled}

\begin{document}

\begin{frontmatter}



\title{PotentRegion4MalDetect: Advanced Features from Potential Malicious Regions for Malware Detection}

\author[inst1]{Rama Krishna Koppanati}

\affiliation[inst1]{organization={Department of Computer Science and Engineering},
            addressline={\\Indian Institute of Technology}, 
            city={Roorkee},
            postcode={247667}, 
            state={Uttarakhand},
            country={India}}

\author[inst2]{Monika Santra}
\author[inst1]{Sateesh Kumar Peddoju}

\affiliation[inst2]{organization={Department of Computer Science and Engineering},
            addressline={\\Pennsylvania State University}, 
            city={State College},
            postcode={PA 16801}, 
            country={USA}}

\begin{abstract}
Malware developers exploit the fact that most detection models focus on the entire binary to extract the feature rather than on the regions of potential maliciousness. Therefore, they reverse engineer a benign binary and inject malicious code into it. This obfuscation technique circumvents the malware detection models and deceives the ML classifiers due to the prevalence of benign features compared to malicious features. However, extracting the features from the potential malicious regions enhances the accuracy and decreases false positives. Hence, we propose a novel model named \textit{PotentRegion4MalDetect} that extracts features from the potential malicious regions. \textit{PotentRegion4MalDetect} determines the nodes with potential maliciousness in the partially preprocessed Control Flow Graph (CFG) using the malicious strings given by \textit{StringSifter}. Then, it extracts advanced features of the identified potential malicious regions alongside the features from the completely preprocessed CFG. The features extracted from the completely preprocessed CFG mitigate obfuscation techniques that attempt to disguise malicious content, such as suspicious strings. The experiments reveal that the \textit{PotentRegion4MalDetect} requires fewer entries to save the features for all binaries than the model focusing on the entire binary, reducing memory overhead, faster computation, and lower storage requirements. These advanced features give an 8.13\% increase in SHapley Additive exPlanations (SHAP) Absolute Mean and a 1.44\% increase in SHAP Beeswarm value compared to those extracted from the entire binary. The advanced features outperform the features extracted from the entire binary by producing more than 99\% accuracy, precision, recall, AUC, F1-score, and 0.064\% FPR. 
\end{abstract}

\begin{keyword}
Advanced Static Malware Analysis, Advanced Features, Control Flow Graph (CFG), Deep Neural Network (DNN), Machine Learning 
\end{keyword}
\end{frontmatter}

\section{Introduction}
\label{section1}
Over 5.4 billion people use the internet as of July 2024, accounting for 67.1\% of the world's population \cite{later1}. However, as internet penetration increases and more users connect online, the threat landscape also expands, inevitably raising the risk of cyberattacks. Norton estimates about 2,200 cyberattacks every day \cite{later3}. Nearly 72\% of businesses experience ransomware attacks as of 2023 \cite{iref1}. AV-TEST Institute records more than 450,000 new malware and potentially unwanted applications (PUA) every day \cite{irefstat}. In 2022 alone, Avast records over 998 million malware, out of which above 76\% of malware targets the Windows Operating System \cite{iref2}. The potential damages caused by unknown malware are so humongous that designing cutting-edge malware detection and prevention methods are essential today.

Malware authors use techniques like malicious code injection, where malicious code is injected inside the benign binary on the selected locations to deceive Machine Learning (ML) models \cite{mld1,mld2}. Though there are many works published to detect malware, most detection approaches extract features from the entire binary instead of focusing on the regions where potent maliciousness is present. As a result, extracted features are more likely to be categorized as benign than malware since they contain more benign features \cite{mld1, mld2}. Further, the approaches that focus on the entire binary to extract features miss the context of the features, leading to less accuracy and more False Positive Rates (FPR). For instance, the approaches that target the entire binary to extract features like the entropy of the binary, histogram of the binary, opcode sequence of the binary, byte sequence of the binary, and API sequences of the binary can be bypassed by tactics such as junk code insertion, equivalent instructions insertion, non-functional instruction insertion, instruction reordering, and dummy API injection. The primary problem with approaches that focus on the entire binary is not understanding why a specific feature appeared at the specific position of the sequences. In other words, they fail to understand the features' context. Another problem is that many approaches focus on shallow features, such as file signature, file header information, import functions, export functions, and digital certificates. Since these features are shallow, sophisticated malware.

Nevertheless, understanding the context of features helps to consider only the significant ones and discard the insignificant ones for building a robust malware detection model. Therefore, a better approach is crucial to identify the potential regions of maliciousness for extracting the robust features, helping to avoid the insignificant features and focusing only on the significant features. This helps to enhance the model's overall performance and decrease the FPR. In this paper, we propose a novel approach that identifies the regions with potential maliciousness and extracts advanced features from the regions. To the best of our knowledge, this is the first research study to concentrate on potential malicious regions and completely preprocessed CFG for feature extraction.

\textit{Motivation:} The distribution of the malicious code is not homogeneous within an executable, and a few portions of the code may exhibit elevated malicious behavior. For instance, the executable may load a legitimate DLL and inject malicious payloads into its memory, embed malicious code in unused sections such as .bss, place it within code caves (unused spaces between instructions), or conceal it between two legitimate code sections. However, focusing on the entire executable file for extracting the features in such scenarios may extract more benign features than malicious features, leading to deceive the ML models. We, therefore, focus on identifying the regions where the potential malicious pieces of code are present and gather features to create an automated malware analysis framework. The following research questions (R) arise in light of the proposed malware analysis technique.

\begin{enumerate}
\item[(R1):] What is the finest way to represent the selected malicious control flow information to prevent contemporary malware evasion techniques, such as dummy code injection?
\item[(R2):] How to detect the potential malicious regions? 
\item[(R3):] How to identify the allowed control flow span around a detected malicious node to create an optimum malware analyzer? 
\item[(R4):] How many are such malicious nodes to consider for a complete feature set generation? 
\item[(R5):] Which features to extract to boost the detection accuracy?
\end{enumerate}

This paper addresses the abovementioned research questions by designing and developing a malware classifier that identifies regions of potential maliciousness within a binary and extracts features from such regions using advanced static analysis.\\
%
%

\textit{Contributions:} The major contributions of the research are as follows:
\begin{itemize}
\item We propose a novel \textit{PotentRegion4MalDetect} model that finds the potential malicious regions in a binary's Control Flow Graph (CFG) and completely preprocessed CFG to extract features. This novel approach allows us to avoid insignificant features and focus only on the significant ones, leading to higher detection accuracy and fewer False Positive Rates (FPR). To our knowledge, no other research has focused on this specific feature extraction method of extracting features from potentially malicious regions, making our model a significant contribution to malware analysis.

\item The CFG of a binary shows all possible paths a binary can take during execution. However, sophisticated malware hides suspicious behavior by obfuscating control flow through tactics such as conditional jumps, loops, and calls. Thus, we preprocess the CFG of a binary to grant immunity from hacker-induced unwanted code without changing the binary's logic and flow.

\item Malware developers use techniques to evade detection, such as junk code insertion, padding, and instructions manipulation. Thus, we capture the structural information of potentially malicious regions in a binary to counter these evasion techniques. This helps to identify behaviorally similar binaries, which exhibit similar patterns of potentially malicious behavior. This way, we can detect and classify typical and zero-day malware.


\item The proposed model enhances data management by minimizing memory usage, increasing computational speed, and reducing storage requirements. It achieves this by generating identical feature entries for binaries that display similar behaviors by concentrating on regions of potential maliciousness and completely preprocessed CFG. This model stands apart from other models that focus on the entire binary for feature extraction.

 \item There is a significant enhancement in performance metrics with advanced features extracted from potential malicious regions and preprocessed CFG rather than the features extracted from the entire binary. It further confirms the effectiveness of the proposed approach for real-time malware classification.
 \end{itemize}

\textit{Organization:} The paper's organization is as follows. Section \ref{section2} discusses the related work in the malware analysis domain,  Section \ref{section3} explains the proposed architecture, and Section \ref{section4} includes the dataset used,  experimental results,  and discussions. Finally, Section \ref{section5} mentions the conclusion.

\section{Related Work}
\label{section2}
Researchers employ static, dynamic, and hybrid analysis to comprehend the characteristics of a binary. Static analysis disassembles the binary without execution, facilitating the extraction of binary inherent features. In contrast, dynamic analysis executes the binary within a controlled environment, enabling the observation of its runtime behavior. Hybrid analysis combines static and dynamic approaches to leverage the advantages of each method, providing a more comprehensive understanding of the binary. Few studies like Pengbin et al. \cite{related2}, Markus et al. \cite{related3}, and Pascal et al. \cite{related4} use dynamic analysis to detect malware. However, they are computationally expensive and vulnerable to adversarial attacks such as anti-VM techniques and Dummy API injection attacks. The works proposed using hybrid analysis, like Asma et al. \cite{related5} and Weijie et al. \cite{ref42}, produce high complexity, high FPR, and inherit the limitations of dynamic analysis. Therefore, we focus on static analysis to extract robust features in this study. Some recent studies on static analysis are as follows:

\textit{Malware Detection Using Opcode Sequences:} Seungho et al. \cite{ref23} use \textit{concolic} execution to generate the largest possible opcode sequence, approximating an actual execution path of malware. They encode sequences of opcodes using opcode-level convolution encoders to train a Recurrent Neural Network (RNN). Although they achieve an accuracy of 96.2\%, their model produces a high FPR even on a small dataset. Abbas et al. \cite{ref11} propose a malware-hunting method combining ML with static analysis. They extract opcode sequences from executables and apply tokenization to convert them into numeric representations. The authors utilize an embedding technique for dimensionality reduction. Then, they feed the processed opcode sequences into a Long Short-Term Memory (LSTM) network for classification. However, they conduct experiments on a small dataset, and their model struggles to understand the semantics and context. Sibel et al. \cite{Sibelg} propose a malware detection model using the opcode sequence of a binary. The authors disassemble the binary to gather the opcodes to construct a directed graph with opcodes as nodes and transition between them as edges. The edge weight is assigned based on the number of times the transition takes between the opcodes. Further, the authors pre-process the constructed directed graph by removing the edges connecting different opcodes. Finally, the authors generate the node degree histogram for the subgraphs and pass these sequences extracted from all the binaries to ML models for classification. Despite the model producing 98\% accuracy, it yields a high FPR. Similarly, Sudan et al. \cite{ref010} also explore static analysis and Neural Networks to design a two-stage malware classifier. After extracting the list of opcodes from the binary executables in the initial stage, the authors employ three different Natural Language Processing (NLP) based techniques separately as vector encoders upon the opcode list. In the second stage, they feed the encoded opcode vectors to the RNN (LSTM cells) to train it as a malware classifier. Despite achieving an accuracy of 91.91\%, the model is computationally complex.

\textit{Malware Detection Using Bytes and Assembly Instructions:} Sota et al. \cite{related1} employ entropy for effective malware detection. The authors calculate the entropy of byte segments in each binary generated using a sliding window approach. Then, the authors apply frequency distribution with interval width `d' to transform the raw entropy values into meaningful feature vectors, capturing the level of randomness. They cluster these feature vectors into low and high-entropy groups using the X-means algorithm and subsequently feed into separate 1D CNN models for classification. Despite the authors achieving an accuracy of 95.1\%, they overlook the contextual information of the binary. As a result, sophisticated malware can manipulate entropy by introducing junk code or adding padding to reduce randomness. Deniz et al. \cite{refy22} investigate the effectiveness of malware detection using Stacked Bidirectional LSTM (BiLSTM) and Generative Pre-trained Transformer 2 (GPT-2) models. They train the models on assembly instructions extracted from PE files' ``.text" section, employing static analysis techniques. They leverage the power of large language models, allowing them to capture both short-term and long-term dependencies while benefiting from static analysis techniques. However, the authors analyze the sequence of instructions without capturing the semantic context of individual instructions. This limitation makes the approach vulnerable to evasion techniques such as instruction reordering, and dummy instruction injection. Moreover, the pre-training of large models required by this method is computationally expensive and time-consuming.

Yifei et al. \cite{Yifeij} propose a model using a binary's bytes and opcode sequences. The authors disassemble the binary and collect the byte and asm files to generate the three $256\times256$ size matrices. They form the first matrix by considering the first 64KB of bytes from a binary to form a $256\times256$ matrix. They form the second matrix from the byte file by removing meaningless hexadecimal bytes repeated continuously five or more times and applying the Continues Bag of Words (CBOW) technique to form a $256\times256$ matrix. Finally, considering the opcodes, they form the third matrix of size $256\times256$. The authors consider these three matrices as three channels in the RGB image and pass it to the neural network model with SEResNet50, Bi-LSTM, and Attention layers. Despite the authors achieving 98.31\% accuracy in malware classification, their model produces less accuracy in malware detection.

\textit{Malware Detection Using API Sequences:} Kohei et al. \cite{Kohei} propose a two-stage malware detection model that leverages API call sequences. The authors collect API sequences from each binary and encode them as integer vectors. In the first stage, each binary's complete API sequence vector is input to a BiLSTM model, which generates a detection score. If this score exceeds a predefined malware threshold, the binary is classified as malware, or it proceeds to check for a benign threshold. If the classification is uncertain, the detection moves to the second stage. In the second stage, the API sequence vector is split into smaller subsequences, each processed individually by the BiLSTM model. If the BiLSTM score exceeds the threshold for each subvector, the binary is immediately classified as malware, and the process stops. If not, the model continues to evaluate subsequent subvectors until a malware score is identified or all subvectors are checked without detecting malware. Although the authors leverage the API sequences to understand the behavior of a binary, their model is threshold-sensitive and vulnerable to adversarial attacks such as dummy API injection attacks \cite{refsyscall}. Hongbi et al. \cite{Hongbi} utilize API sequences along with other shallow features, including MajorLinkVersion, CheckSum, DllCharacteristics, and ImageBase, for malware detection. The API sequences extracted from binaries are hashed to values ranging from 0 to 255, resulting in 256 features. The authors subsequently pass these features, along with other shallow features, through autoencoders to reduce the dimensionality and enhance the detection process's performance and efficiency. While the authors attained an accuracy of 94.24\%, their use of hashing to map all APIs to values ranging from 0 to 255 introduces the risk of collisions, which could result in the loss of critical information. Vidhi et al. \cite{Vidhi} employ API calls for malware detection. The authors extract API calls from all binaries in the dataset and identify the unique APIs based on their presence across the binaries. They then generate a one-hot encoded vector for each binary to represent the presence or absence of each unique API. The vectors undergo feature selection utilizing a Random Forest method for identifying 45 essential features. The authors subsequently pass the selected features to various ML models for classification. Despite the authors achieving an accuracy of 92.67\%, their model is vulnerable to dummy API injection attacks \cite{refsyscall}.
\begin{table}[!b]
\renewcommand{\tabcolsep}{1.5pt}
\caption{PotentRegion4MalDetect Vs. SOTA Static Malware Detection Models.}
\scriptsize
\begin{tabular}{|l|l|c|c|c|c|c|c|c|}
\hline
\multicolumn{1}{|c|}{\textbf{Approach}}                                                                                                                                                                                       & \multicolumn{1}{c|}{\textbf{Approach}}                                                                                                                                               & \textbf{CA}  & \textbf{FPMR} & \textbf{RJCP} & \textbf{REI} & \textbf{RNFI} & \textbf{RIR} & \textbf{RDAPII} \\ \hline
\begin{tabular}[c]{@{}l@{}}Seungho et al. \cite{ref23}\\ Abbas et al. \cite{ref11}\\ Sibel et al. \cite{Sibelg}\\ Sudan et al. \cite{ref010}\end{tabular} & Opcode sequence                                                                                                                                                                      & \xmark           & \xmark            & \xmark            & \xmark           & \xmark            & \xmark           & -               \\ \hline
\begin{tabular}[c]{@{}l@{}}Sota et al. \cite{related1}\\ Yifei et al. \cite{Yifeij}\end{tabular}                                                                                            & Bytes                                                                                                                                                                                & \xmark           & \xmark            & \xmark            & \xmark           & \xmark            & \xmark           & -               \\ \hline
Deniz et al. \cite{refy22}                                                                                                                                                                                   & \begin{tabular}[c]{@{}l@{}}Assembly instructions\end{tabular}                                                                                                                      & \xmark           & \xmark            & \xmark            & \xmark           & \xmark            & \xmark           & -               \\ \hline
\begin{tabular}[c]{@{}l@{}}Kohei et al. \cite{Kohei}\\ Hongbi et al. \cite{Hongbi}\\ Vidhi et al. \cite{Vidhi}\end{tabular}                                                & API sequence                                                                                                                                                                         & \xmark           & \xmark            & -             & -            & -             & -            & \xmark              \\ \hline
\textbf{PotentRegion4MalDetect}                                                                                                                                                                                                    & \textbf{\begin{tabular}[c]{@{}l@{}}Advanced Features:\\ opcode sequence,\\ API sequences, Subgraph\\ Signature, Trigrams, NOP\\ count, and Section ratio.\end{tabular}} & \textbf{\cmark} & \textbf{\cmark}  & \textbf{\cmark}  & \textbf{\cmark} & \textbf{\cmark}  & \textbf{\cmark} & \textbf{\cmark}                                     \\ \hline
\multicolumn{9}{c}{\begin{tabular}[c]{@{}c@{}}\tiny{CA: Context-Aware, FPMR: Features from Potential Malicious Regions, RJCP: Resistance to Junk code and Padding,}\\ \tiny{REI: Resistance to Equivalent Instructions, RNFI: Resistance to Non-Functional Instructions, RIR: Resistance to}\\ \tiny{Instructions Reordering, RDAPII: Resistance to Dummy API  Injection}\end{tabular}}
\end{tabular}
\label{relatedtable1}
\end{table}
\begin{table}[t]
\renewcommand{\tabcolsep}{1.5pt}
\caption{PotentRegion4MalDetect Vs. SOTA Dynamic Malware Detection Models.}
\scriptsize
\begin{tabular}{|l|l|c|c|c|c|c|c|c|}
\hline
\multicolumn{1}{|c|}{\textbf{Approach}}     & \multicolumn{1}{c|}{\textbf{Feature Used}}                                                                                                                                                  & \textbf{\begin{tabular}[c]{@{}c@{}}CA\end{tabular}} & \textbf{\begin{tabular}[c]{@{}c@{}}FPMR\end{tabular}} & \textbf{\begin{tabular}[c]{@{}c@{}}RJCP\end{tabular}} & \textbf{\begin{tabular}[c]{@{}c@{}}REI\end{tabular}} & \textbf{\begin{tabular}[c]{@{}c@{}}RNFI\end{tabular}} & \textbf{\begin{tabular}[c]{@{}c@{}}RIR\end{tabular}} & \textbf{\begin{tabular}[c]{@{}c@{}}RDAPII\end{tabular}} \\ \hline
Pengbin et al. \cite{related2}                & API sequences                                                                                                                                                                      & \xmark                                                               & \xmark                                                                                                      & \cmark                                                                                        & \cmark                                                                                          & \cmark                                                                                              & \cmark                                                                                          & \xmark                                                                                      \\ \hline
Markus et al. \cite{related3}                 & \begin{tabular}[c]{@{}l@{}}Action, target, name, process\\ path, process ending, target\\ path, and target ending\end{tabular}                                              & \xmark                                                               & \xmark                                                                                                      & \cmark                                                                                        & \cmark                                                                                          & \cmark                                                                                              & \cmark                                                                                          & \cmark                                                                                     \\ \hline
Pascal et al. \cite{related4}                 & \begin{tabular}[c]{@{}l@{}}Features from memory dumps\end{tabular}                                                                                                               & \xmark                                                               & \xmark                                                                                                      & \cmark                                                                                        & \cmark                                                                                          & \cmark                                                                                              & \cmark                                                                                          & -                                                                                       \\ \hline
\textbf{PotentRegion4MalDetect}                       & \textbf{\begin{tabular}[c]{@{}l@{}}Advanced Features:\\ opcode sequence,\\API sequences, Subgraph\\Signature, Trigrams, NOP\\count, and Section ratio.\end{tabular}} & \textbf{\cmark}                                                     & \textbf{\cmark}                                                                                            & \textbf{\cmark}                                                                               & \textbf{\cmark}                                                                                 & \textbf{\cmark}                                                                                     & \textbf{\cmark}                                                                                 & \textbf{\cmark}                                                                            \\ \hline
\multicolumn{9}{c}{\begin{tabular}[c]{@{}c@{}}\tiny{CA: Context-Aware, FPMR: Features from Potential Malicious Regions, RJCP: Resistance to Junk code and Padding,}\\ \tiny{REI: Resistance to Equivalent Instructions, RNFI: Resistance to Non-Functional Instructions, RIR: Resistance to}\\ \tiny{Instructions Reordering, RDAPII: Resistance to Dummy API  Injection}\end{tabular}}
\end{tabular}
\label{relatedtable2}
\end{table}
\begin{table}[t]
\renewcommand{\tabcolsep}{1.5pt}
\caption{PotentRegion4MalDetect Vs. SOTA Hybrid Malware Detection Models.}
\scriptsize
\begin{tabular}{|l|l|c|c|c|c|c|c|c|}
\hline
\multicolumn{1}{|c|}{\textbf{Approach}}     & \multicolumn{1}{c|}{\textbf{Feature Used}}                                                                                                                                                  & \textbf{\begin{tabular}[c]{@{}c@{}}CA\end{tabular}} & \textbf{\begin{tabular}[c]{@{}c@{}}FPMR\end{tabular}} & \textbf{\begin{tabular}[c]{@{}c@{}}RJCP\end{tabular}} & \textbf{\begin{tabular}[c]{@{}c@{}}REI\end{tabular}} & \textbf{\begin{tabular}[c]{@{}c@{}}RNFI\end{tabular}} & \textbf{\begin{tabular}[c]{@{}c@{}}RIR\end{tabular}} & \textbf{\begin{tabular}[c]{@{}c@{}}RDAPII\end{tabular}} \\ \hline
Asma et al. \cite{related5}                       & API sequences                                                                                                                                                                               & \xmark                                                               & \xmark                                                                                                      & \cmark                                                                                        & \cmark                                                                                          & \cmark                                                                                            & \cmark                                                                                        & \xmark                                                                                      \\ \hline
Weijie et al. \cite{ref42} & \begin{tabular}[c]{@{}l@{}}PE section information, API\\and DLL sequences, and\\information about files,\\ networks, and registries\end{tabular}                                    & \xmark                                                               & \xmark                                                                                                      & \cmark                                                                                        & \cmark                                                                                          & \cmark                                                                                            & \cmark                                                                                        & \xmark                                                                                      \\ \hline
\textbf{PotentRegion4MalDetect}                       & \textbf{\begin{tabular}[c]{@{}l@{}}Advanced Features:\\ opcode sequence,\\API sequences, Subgraph\\Signature, Trigrams, NOP\\count, and Section ratio.\end{tabular}} & \textbf{\cmark}                                                     & \textbf{\cmark}                                                                                            & \textbf{\cmark}                                                                               & \textbf{\cmark}                                                                                 & \textbf{\cmark}                                                                                     & \textbf{\cmark}                                                                                 & \textbf{\cmark}                                                                            \\ \hline
\multicolumn{9}{c}{\begin{tabular}[c]{@{}c@{}}\tiny{CA: Context-Aware, FPMR: Features from Potential Malicious Regions, RJCP: Resistance to Junk code and Padding,}\\ \tiny{REI: Resistance to Equivalent Instructions, RNFI: Resistance to Non-Functional Instructions, RIR: Resistance to}\\ \tiny{Instructions Reordering, RDAPII: Resistance to Dummy API  Injection}\end{tabular}}
\end{tabular}
\label{relatedtable3}
\end{table}

\textit{Key Considerations:} Most of the works proposed in the literature focus on opcode sequences, bytes, assembly instructions, and API sequences of the entire binary. Since the malicious code is not homogenously distributed throughout the binary, these approaches are vulnerable to evasion techniques such as replacing instructions with equivalent instructions, altering the program's control flow, adding non-functional instructions, changing the order of independent instructions, and dummy API injection attacks. On the other hand, the dynamic and hybrid analysis models require more computation and produce more FPR. Therefore, in the proposed model, we extract features from the potential malicious regions and completely preprocess CFG with context awareness of the features in malware detection. Further, the proposed model resists all the above-listed evasion techniques and achieves higher accuracy and lower FPR than the SOTA models due to its ability to extract features from regions of importance. The comparison of the proposed work with the SOTA static, dynamic, and hybrid malware detection works regarding their resistance to various attacks and context-aware feature extraction is shown in Table \ref{relatedtable1}, Table \ref{relatedtable2}, and Table \ref{relatedtable3}, respectively.

\section{Proposed PotentRegion4MalDetect Model}
\label{section3}
\begin{figure}[!t]
\centering
\includegraphics[width=\textwidth, trim = 0.25cm 6.75cm 0cm 6.5cm, clip]{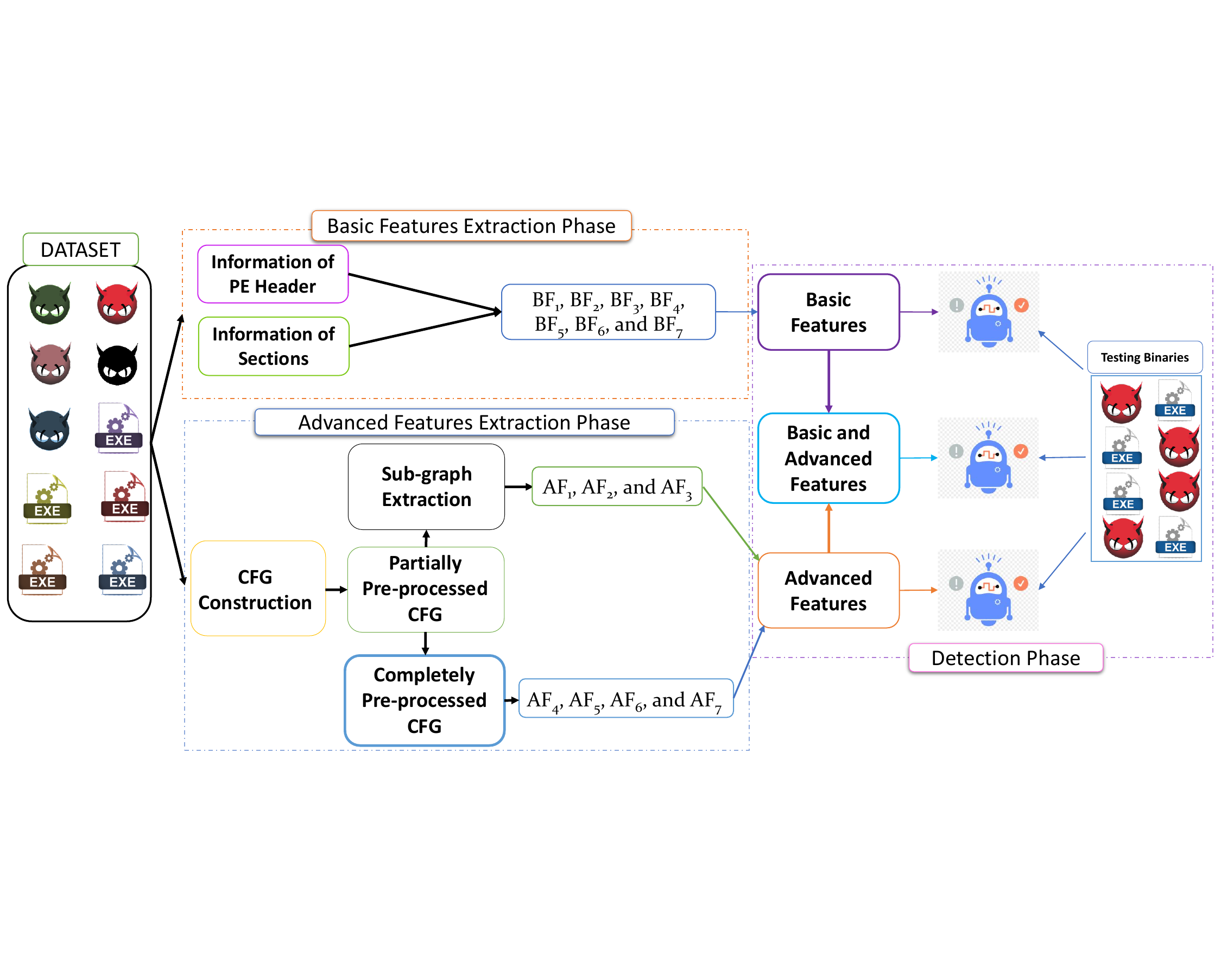}
\caption{Complete PotentRegion4MalDetect Framework}
\label{fig7}
\end{figure}

We propose a novel model named `\textit{PotentRegion4MalDetect}' to detect malware using advanced static analysis. This section deals with designing and implementing the framework of the proposed \textit{PotentRegion4MalDetect} model. The framework comprises two phases: Feature Extraction and ML classifier implementation. The feature extraction phase includes advanced feature extraction, which uses the advanced static analysis approach to extract the features from regions of potential maliciousness and completely preprocessed CFG. We illustrate the design of the feature extractor followed by the complete proposed approach. The complete \textit{PotentRegion4MalDetect} framework and module decomposition are shown in Figure \ref{fig7}, where each phase is represented in a different color bar at the bottom. We discuss each phase of the proposed framework in detail in the following sub-sections.

\subsection{Feature Extraction}
\label{section31}
We explain the process of employing advanced static analysis techniques to identify the potential malicious regions of the CFG and completely preprocessing the CFG for advanced feature generation. 
To simplify things, we refer to the shallow features extracted from the whole binary as \textit{``basic features"} and the other type as \textit{``advanced features."}
\subsection{Advanced Feature Extraction}
\label{section312}
Code injection is a powerful and widely used technique in malware development, where malware developers inject malicious code inside benign binary. Whereas the location to inject malicious code inside benign binary can be optimum selected or random depending on the skills of a malware developer \cite{mld2}. This technique helps bypass security measures and decreases the chances of finding malware by fooling ML classifiers \cite{mld1,mld2}. Malware analysts frequently extract features from the binary as a whole rather than concentrating on the areas where malicious code is injected. As a result, extracted features are more likely to be categorized as benign than malware since they contain more benign features \cite{mld1,mld2}. \\
However, the question arises of how to find the regions inside a binary with the most potential maliciousness. We address this question with the help of the following Subsection, which includes two steps. In the Initial step, we construct a Control Flow Graph (CFG) from the disassembled binary and pre-process the constructed CFG to prevent specific malware evasion strategies. In the second step, we extract potential malicious regions in the form of sub-graphs from the partially pre-processed CFG. The following sections discuss the complete operation of extracting advanced features from potential malicious regions and completely pre-processed CFGs.\\
\textit{Remark: Most of the proposed works focus on the entire binary to extract features to identify the category of a binary. In contrast, the proposed work focuses on regions with the most potential maliciousness and extracts critical features. The more focused features help decrease the FPR and achieve heightened results.}
\subsection{Modified Control Flow Graph (CFG)}
\label{para1}
\begin{figure}[!t]
\centering
\includegraphics[width=0.8\textwidth, trim = 2.5cm 1cm 2.6cm 0.6cm, clip]{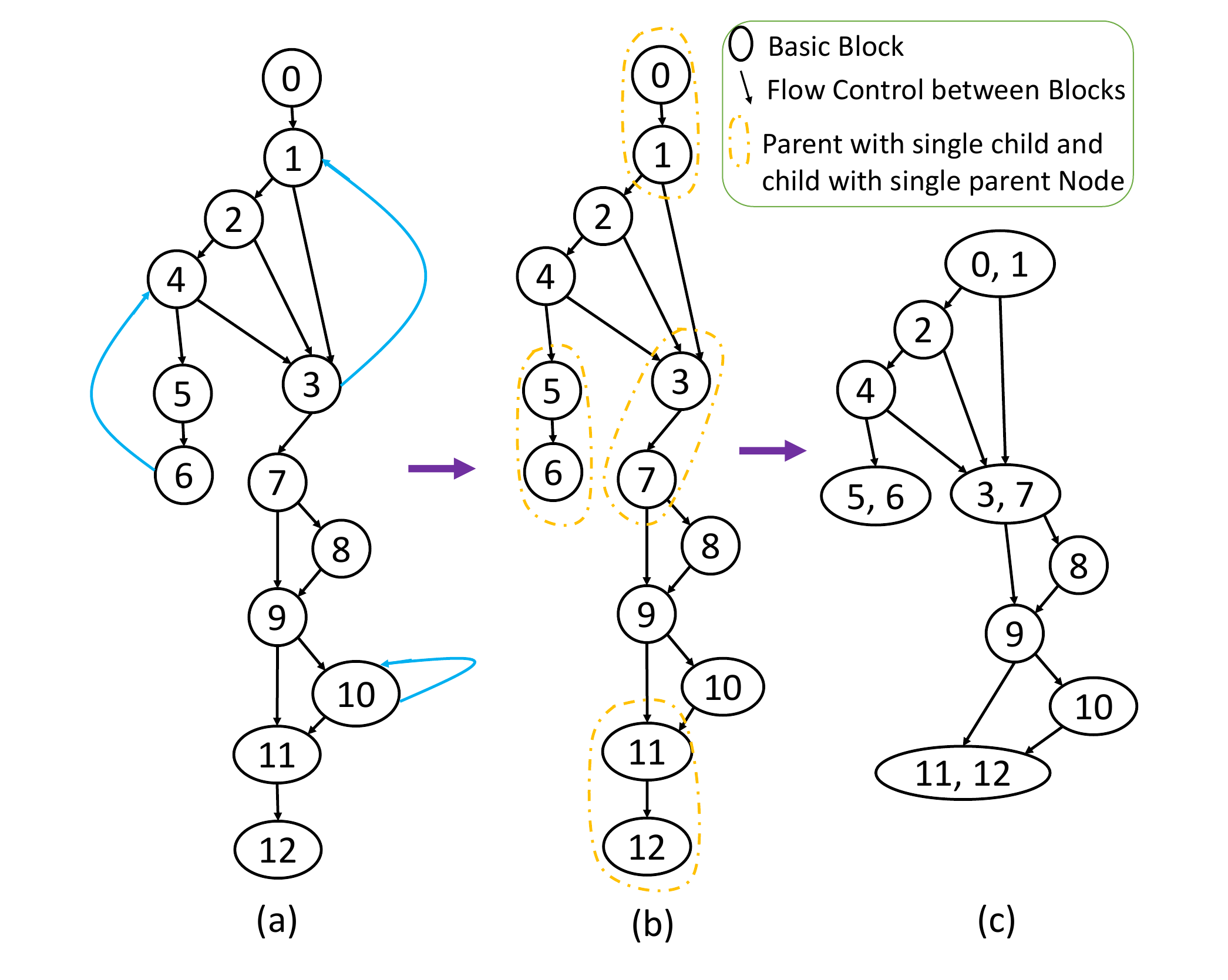}
\caption{Modified CFG: (a) Original CFG, (b) Partially Pre-processed CFG, and where the loops are removed 
(c) Completely Pre-processed CFG where the nodes with the single child are merged with its parent node of the partially pre-processed CFG. (nodes 0,  3,  5,  and 11 have single nodes, as shown in the middle tree. After pre-processing the CFG,  nodes 0 and 1,  nodes 5 and 6,  nodes 3 and 7,  and node 11 and 12 are combined as a single node)
}
\label{fig3}
\end{figure}

This module takes the disassembled binary information as input. By parsing the input information, it generates the CFG by identifying the basic blocks. However, attackers inject dummy code to modify the CFG without changing the code's intended flow to evade the existing signature-based detection approaches. In this scenario, using the constructed CFG alone may not increase the detection accuracy. Hence, we pre-process the originally constructed CFG. This involves removing any loops and merging a parent node with its child node when the parent has only one child, and the child has only one parent. This approach ensures that the intended flow of the program is preserved. This is important as the generated CFG is immune to any small changes in the CFG. 
Figure \ref{fig3} depicts a sample of the CFG pre-processing in action. We start by removing all the loops. Then, if any parent node has a single child, we merge that child node with its parent since it does not affect the logic or flow of the program. For instance, in Figure \ref{fig3}, node 0 merges with its child node 1, node 5 with its child 6, node 3 with 7, and node 11 with 12, as they have a single child node. This answers the research question (R1).    
\subsection{Ranking Malicious Nodes in a CFG}
\label{para2}
 We use an open-source framework called StringSifter \cite{ref10}, developed by Mandiant, for generating the list of malicious strings with their associated maliciousness scores. StringSifter \cite{ref10} is a powerful ML tool that automatically ranks malicious strings based on their relevance for malware analysis. We use this tool on every binary to identify the top malicious strings based on score. Using those malicious strings, we identify the potential malicious nodes in a CFG in the subsequent steps. We deploy this method by analyzing the fact that the nodes with malicious strings, along with their surrounding nodes, can increase detection accuracy. This answers the research question (R2). Some malicious strings and associated maliciousness scores are shown in Table \ref{table3}.
 \begin{table}[t]
\centering
\scriptsize
\caption{\normalsize{Sample Strings and Their Maliciousness Scores}}
\begin{tabular}{|l|c|}
\hline
\multicolumn{1}{|c|}{\textbf{String}}                                                                     & \textbf{\begin{tabular}[c]{@{}c@{}}Maliciousness Score\end{tabular}} \\ \hline
Vmx32to6.exe                                  & 9.90                                                                    \\ \hline
CONNECT \%s:\%i HTTP/1.0                        & 9.57                                                                    \\ \hline
SOFTWARE\textbackslash Microsoft\textbackslash Windows\textbackslash CurrentVersion\textbackslash Run & 9.56                                                                    \\ \hline
StubPath                                      & 7.60                                                                    \\ \hline
\end{tabular}
\label{table3}
\end{table}
\begin{figure}[!t]
\centering
\includegraphics[width=0.5\textwidth, trim = 0.75cm 1cm 0.65cm 0.3cm, clip]{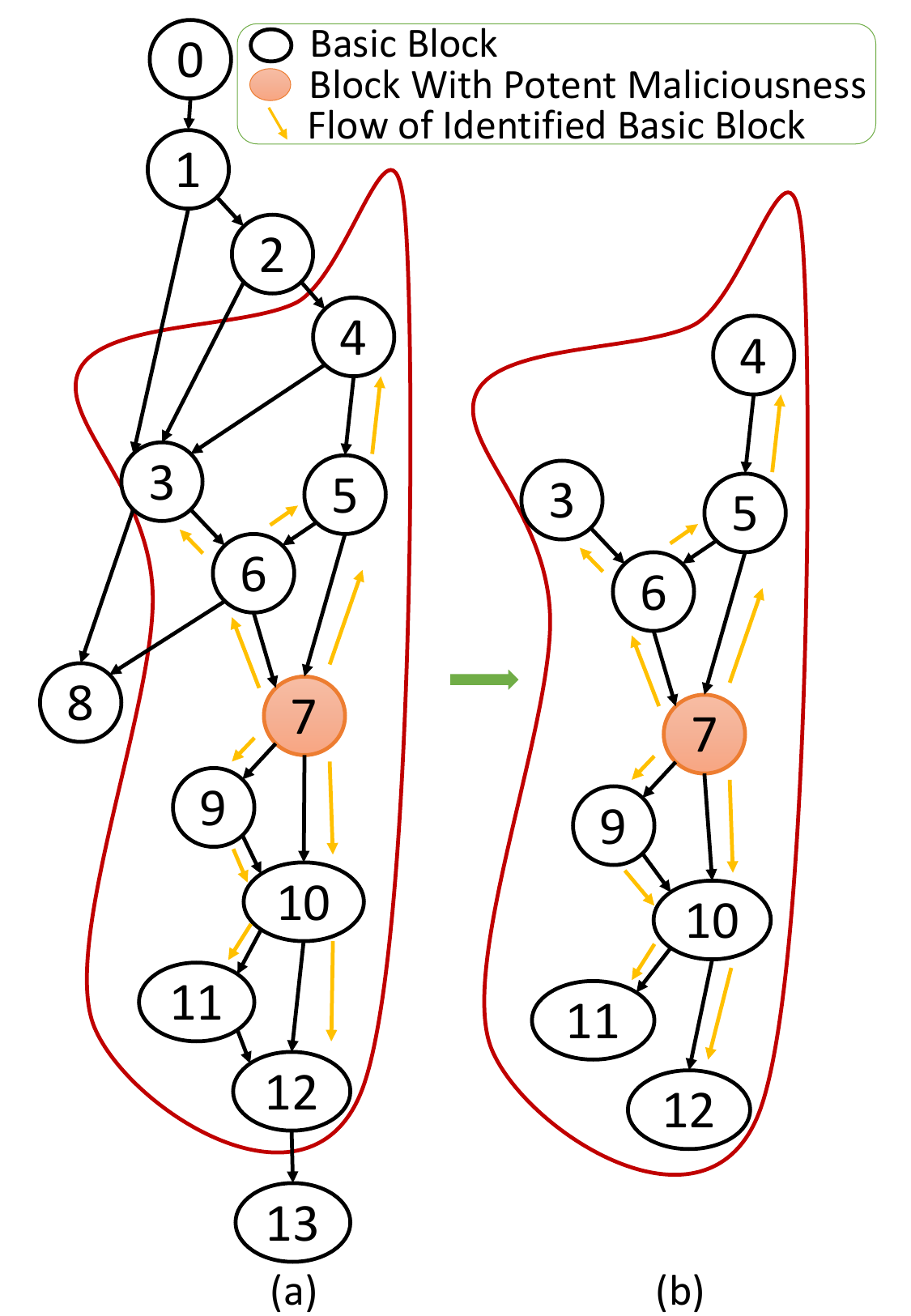}
\caption{Subgraph Extraction from CFG of the Binary: (a) Partially Pre-processed CFG (b) Subgraph of the CFG (visiting the top and bottom two-level nodes of the identified node, i.e.,  7 with top malicious string score).
}
\label{fig4}
\end{figure} 
 \subsection{Sub-graph Extraction}
\label{para33}
After identifying the nodes with top-score malicious strings,  we extract subgraphs consisting of some surrounding nodes, which are at most two levels away from the identified malicious node in the CFG, as portrayed in Figure  \ref{fig4}. We take the number of levels surrounding the selected malicious node as `two' based on our empirical evidence of increased accuracy with the taken dataset as shown in Figure \ref{table12}. We keep these parameters configurable for the later tuning process. For instance,  in Figure \ref{fig4},  we select node seven as the potential malicious node considering the presence of the top score malicious string in it. We extract the subgraph with top and bottom two-level nodes surrounded by node 7 (i.e., 5,  6,  4,  3,  9,  10,  11,  12). This answers the research question (R3).  
\subsection{Advanced Feature Extraction from Sub-graphs\label{para3}} This section explains the advanced features using sub-graphs in detail. The selected features from the sub-graphs are BFS-based opcodes,  API sequences,  and signatures, 
as shown in Figure \ref{fig5}.
\begin{itemize}
\item[-] \textit{Opcode Sequence Generation:} Opcodes or operational codes are a basic block of computing instructions. Previous research indicates that the sequence of opcodes is usable as an efficient predictor for malware detection \cite{ref78}. Rather than considering all opcodes from the whole binary, we focus on the sequence of opcodes from a targeted part, i.e., from the sub-graph of a binary. It is essential to consider the flow of opcodes in the sub-graph as there is less probability for attacker-induced dummy instructions to present surrounding such malicious nodes. 
\item[-] \textit{API Sequence Generation:} As APIs contribute actively to the malware detection process  \cite{ref13, ref30}, we gather the imported API call sequences to approximate the run time behavior of a binary.  We traverse each node of the extracted sub-graph in a BFS fashion and create a sequence of APIs list as they come across in the traversed path.  We use the API sequence based on the fact that,  even if the attackers insert dummy instructions to evade opcode-based or signature-based detection,  there is a high probability that they are detected. The mapping of APIs to the CFG nodes is explained in detail in Subsection \ref{api_mapper}.

\begin{figure}[!t]
\centering
\includegraphics[width=0.8\textwidth, trim = 4.5cm 0.5cm 2.4cm 0.5cm, clip]{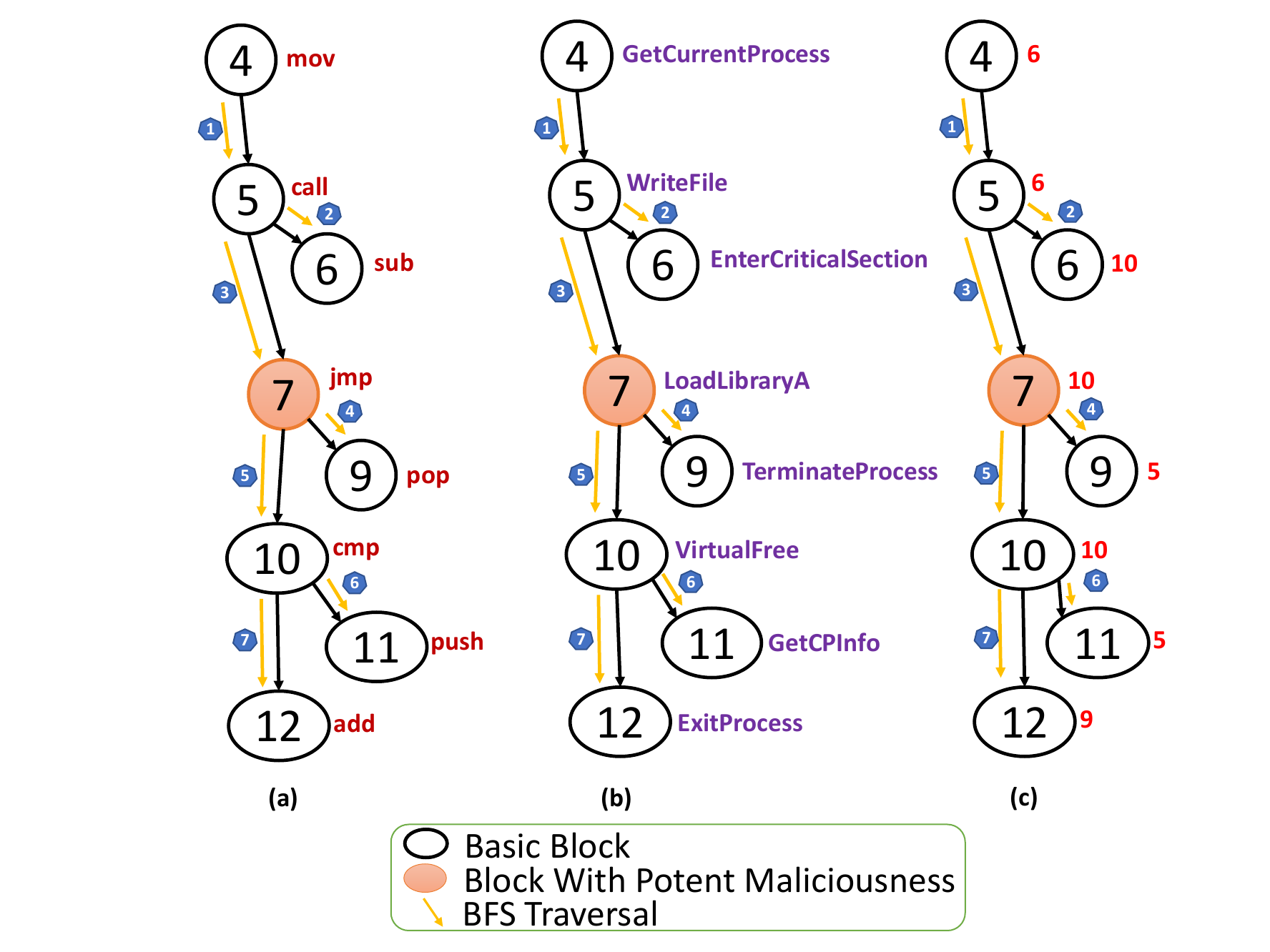}
\caption{Subgraph Feature Extraction: (a) Opcode Sequence in the BFS traversal order: mov,  call,  sub,  jmp,  pop,  cmp,  push,  add 
(b) API Sequence in the BFS traversal order: GetCurrentProcess,  WriteFile,  EnterCriticalSection,  LoadLibraryA, TerminateProcess,  VirtualFree,  GetCPInfo,  ExitProcess 
(c) Signature sequences in the BFS traversal order: 6,  6,  10,  10,  5,  10,  5,  9
}
\label{fig5}
\end{figure}
\item[-] \textit{Signature Generation:\label{sign_ref}} We generate a signature to identify any subgraph with a similar structure. The signature generation of the subgraph is as follows. Firstly,  we assign each node a unique number. We generate the assigned number using two parameters as mentioned in \cite{ref29}. The least significant 2 bits of a number indicate the number of children,  and the remaining 6 bits represent the number of parent nodes that node has \cite{ref29}. Figure \ref{fig5} shows an example of signature generation. Following this approach, node 5 gets number 6 in Figure \ref{fig5}, which is generated by representing the least significant 2 bits by its number of children,  i.e.,  2, and the remaining 6 bits indicate the number of parents,  i.e., 1, as shown in Figure \ref{fig6}. We generate the signature considering that even if a few minor parts of the CFG change, the whole signature does not vary. The extracted signature value from each subgraph is reshaped into a vector of size 100 to ensure consistency.
\begin{figure}[]

\centering
\includegraphics[width=0.85\textwidth, trim = 3cm 5cm 2.5cm 4cm, clip]{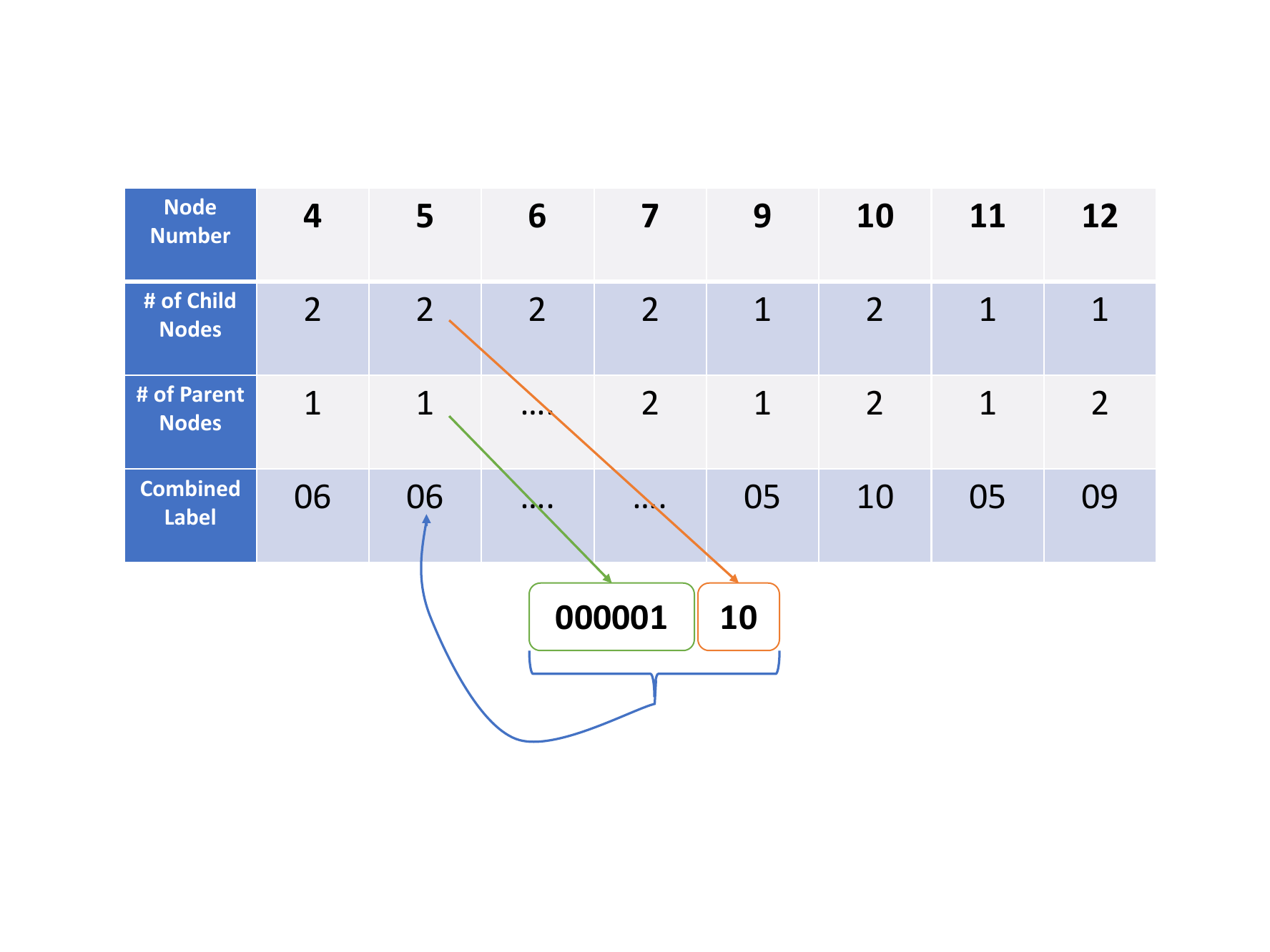}
\caption{Signature Generation Scheme}
\label{fig6}
\end{figure}
\end{itemize}
\label{para5}
We select the top ten potential malicious sub-graphs for advanced feature extraction. We select the number `ten' based on our empirical observations, as shown in Figure \ref{table12}. We encounter four cases while trying to identify the top ten potential malicious nodes in the CFG to create sub-graphs.  In the first case, We find ten or more malicious nodes in the partially pre-processed CFG. However, from the discovered malicious nodes, we select ten nodes as our framework considers the top ten potential malicious nodes. In the second case, we find greater than zero or less than ten potential malicious nodes. In such situations,  we select the originally detected nodes for feature generation and fill the remaining node features with default values to maintain the standard dimension. In the third case, we find no potential malicious nodes in the CFG, as it contains zero malicious strings. In that scenario,  we take only one node,  i.e.,  either the first node in the CFG or one of the middle nodes in the CFG. We fill the rest of the remaining nine node features with default values. Finally, the last case considers either the CFG construction from a binary is unsuccessful or an exception handler is triggered in the framework. In that scenario, we fill advanced features with default values.
\subsection{Advanced Features Extraction from Completely Pre-processed CFG\label{para4}}Obfuscation techniques, such as packing and encryption, transform the appearance of a binary to hinder analysis and conceal malicious content, such as suspicious strings, making detection significantly challenging. However, in this work, apart from extracting the features from the potential malicious regions, we also extract features from the completely pre-processed CFG. The features extracted from the complete pre-processed CFG are as follows.
\begin{itemize}
\item \textit{Opcode Trigram:} N-grams are a combination of adjacent words or letters of length `N,' where N is a positive integer. Many studies use N-grams to identify malware,  where N-grams represent the features, such as System calls,  APIs, and Opcodes. In our work, we use Term Frequency Inverse Document Frequency (TF-IDF) for collecting the top fifteen opcode trigrams from a binary. We use the TF-IDF technique to weigh a trigram within the content to know the importance of that specific opcode trigram based on counting the number of times it is mentioned in the content. We select the number `fifteen' by analyzing the outcomes of the experiments, as shown in Figure \ref{table12}. 
\item \textit{Signature of the Complete CFG:} We also generate the signature for the completely pre-processed CFG. This feature matches any binary concerning its control flow logic similarity. We compute the signature as defined in \ref{sign_ref}, reshape it into a consistent 200-dimensional vector, and proceed with normalization.

\item \textit{Number of NOP Operations:} The number of NOP operations is significant for detecting the amount of dummy code within a binary. The high NOP operation count indicates a higher probability of the binary being malicious. We extract the NOP count from the whole binary.
\item \textit{Section Ratio:} The experiments show that the section ratio value for some malware is greater than 1.5 \cite{refpractice}. Thus, it is a significant feature to distinguish the malware from the benign. As shown in Equation \ref{eq1},  we store the section ratio value as 1 in the feature set if any section ratio is greater than 1.5,  else zero. We select this value as 1.5 depending on the empirical evidence of increased accuracy with the taken dataset as shown in Figure \ref{table12}.
\end{itemize}
\begin{small}
\begin{equation}
Section\ Ratio= 
\begin{cases}
    1,& \text{if } (\exists(\frac{Section\ Virtual\ Size}{Section\ Physical\ Size}))>1.5\\
    0,              & \text{otherwise}
\end{cases}
\label{eq1}
\end{equation}
\end{small}
The comprehensive list of PotentRegion4MalDetect features, along with their corresponding dimensions and details of the basic features extracted from the entire binary, are presented in Table \ref{features_details}. The Algorithm \ref{algo1} illustrates each of the steps, and the complete PotentRegion4MalDetect architecture is shown in Figure \ref{fig7}.
\begin{table}[]
\caption{Details of PotentRegion4MalDetect Features.}
\centering
\label{features_details}
\begin{tabular}{|lc|c|}
\hline
\multicolumn{2}{|c|}{\textbf{Advanced Features}}                                                                                                                                           & \textbf{Dimension} \\ \hline
\multicolumn{1}{|l|}{\begin{tabular}[c]{@{}l@{}}Ten Subgraphs\\ API sequences\end{tabular}}    & \multirow{3}{*}{\begin{tabular}[c]{@{}c@{}}Potential\\ Malicious\\ Regions\end{tabular}}  & 100                \\ \cline{1-1} \cline{3-3} 
\multicolumn{1}{|l|}{\begin{tabular}[c]{@{}l@{}}Ten Subgraphs\\ Opcode sequences\end{tabular}} &                                                                                           & 100                \\ \cline{1-1} \cline{3-3} 
\multicolumn{1}{|l|}{\begin{tabular}[c]{@{}l@{}}Ten Subgraphs\\ Signatures\end{tabular}}       &                                                                                           & $10\times 100$           \\ \hline
\multicolumn{1}{|l|}{\begin{tabular}[c]{@{}l@{}}Whole CFG\\ Signature\end{tabular}}            & \multirow{4}{*}{\begin{tabular}[c]{@{}c@{}}Completely\\ Pre-processed\\ CFG\end{tabular}} & 200                \\ \cline{1-1} \cline{3-3} 
\multicolumn{1}{|l|}{Fifteen Trigrams}                                                         &                                                                                           & 20                 \\ \cline{1-1} \cline{3-3} 
\multicolumn{1}{|l|}{\begin{tabular}[c]{@{}l@{}}Number of\\ NOPs\end{tabular}}                 &                                                                                           & 1                  \\ \cline{1-1} \cline{3-3} 
\multicolumn{1}{|l|}{\begin{tabular}[c]{@{}l@{}}Ratio of\\ virtual size\end{tabular}}          &                                                                                           & 1                  \\ \hline
\multicolumn{2}{|c|}{\textbf{Total}}                                                                                                                                                       & 1422               \\ \hline
\end{tabular}
\end{table}

\begin{algorithm}[!b]
\DontPrintSemicolon
  \scriptsize{\KwInput{Set of Binaries ($B_1,B_2,….,B_n$) to be tested }}
  \scriptsize{\KwOutput{All the extracted Advanced features}}
\tcc{\scriptsize{`\textit{n}' malicious node extraction from CFG}}
 \scriptsize{Apply $radare2(B_i)$ to extract CFG in JSON format}
 
 \scriptsize{$Mal\_strings\_score = StringSifter(B_i)$} 
 
 \scriptsize{$ Nodes =Node\_Finder(Mal\_strings\_score)$}  
 
\For{ \scriptsize{j = 0 to len(nodes)}}
{
\tcc{\scriptsize{Subgraph Extraction for each node}}
\scriptsize{$Subgraph = Extract\_Subgraph(CFG, Node_j)$} 

\scriptsize{$Subgraph\_BFS = BFS(Subgraph)$}

\For{ \scriptsize{k = 0 to len(Subgraph\_BFS)}}
{
\tcc{\scriptsize{Opcode, API, and signature extraction}}
\scriptsize{$Node\_Opcode\_Sequence.append$
$(Opcode\_sequence(Subgraph\_BFS\_node_k))$}

\scriptsize{$Node\_API\_Sequence.append$
$(API\_sequence(Subgraph\_BFS\_node_k))$}

\scriptsize{$Node\_Sign.append$
$(Node\_Signature(Subgraph\_BFS\_node_k))$}
}
\tcc{\scriptsize{Store the extracted opcode, API, and signature of all extracted subgraphs}}
\scriptsize{$Subgraph\_opcode\_sequence.append$
$(Node\_Opcode\_Sequence)$}

\scriptsize{$Subgraph\_API\_sequence.append$
$(Node\_API\_Sequence)$}

\scriptsize{$Subgraph\_Sign\_sequence.append$
$(Node\_Sign)$}
}
\tcc{\scriptsize{Signature, NOP count, and ten trigrams from the pre-processed CFG}}
\scriptsize{$CFG\_Approx\_Sign = Signature(Pre-processed\ CFG)$}

\scriptsize{$NOP\_count = NOP\_Count(CFG)$}

\scriptsize{$Trigram = Trigram\_CFG(CFG)$}

\tcc{\scriptsize{Section ratio value definer}}
\If{\scriptsize{$\exists(\frac{physical\_section\_size}{virtual\_sectio\_size})>1.5$}}
{
\scriptsize{$Section\_Ratio = 1$}
}
\Else
{
\scriptsize{$Section\_Ratio = 0$}
}

\caption{Advanced Feature Extraction ($B_i$)}
\label{algo1}
\end{algorithm}
\subsection{Case Study} 
Consider two Spywares with MD5 hashes \textit{420b0e81c7588b5f9c5e13983692-abe4} and \textit{e302e9f4029045f750cb6314ffad1488} for the demonstration. The first Spyware has 391 lines of code, and the second Spyware has 3448 lines of code. However, the first binary has significantly fewer lines of code and exhibits malware behavior throughout. On the other hand, the second binary has more lines of code, and there are some specific regions where the most potential maliciousness is present. If we apply traditional methods to the second binary, there is a chance that the features will include even the benign features and may bypass ML models. Nevertheless, the proposed PotentRegion4MalDetect targets the potential malicious regions and extracts the features from those regions. The proposed model identifies CFG nodes with addresses [`\textit{0x10004e6f}', `\textit{0x10004ef8}', `\textit{0x10004f09}', `\textit{0x10004ee6}', `\textit{0x10004f31}', `\textit{0x10 004f1e}', `\textit{0x10004ed4}'] as malicious nodes covering the entire CFG of the first binary. On the other hand, it identifies the nodes with addresses [`\textit{0x4032e2}', `\textit{0x4033f4}', `\textit{0x4030cb}', `\textit{0x403272}', `\textit{0x403239}', `\textit{0x4032a7}', `\textit{0x403442}', `\textit{0x4 032b7}', `\textit{0x40342d}', `\textit{0x40327e}'] as malicious nodes in the second binary targeting only the potential malicious regions without including the unwanted features.
\subsection{Map APIs to the CFG Nodes}
\label{api_mapper}
Mapping the APIs in the binary to the nodes in the CFG is a challenging problem, as nodes in the CFG may have many recursive function calls. So, the available disassemblers, such as radare2 and Ghidra, do not have the functionality to map the binary's APIs to the CFG nodes. Therefore, we use the mechanism presented by us in \cite{msg_rama} to map the APIs to those in the CFG, as it is an essential part of the proposed work. 

\begin{figure}[!t]
\centering
\includegraphics[width=\textwidth, trim = 0.7cm 0.4cm 0.9cm 0.35cm, clip]{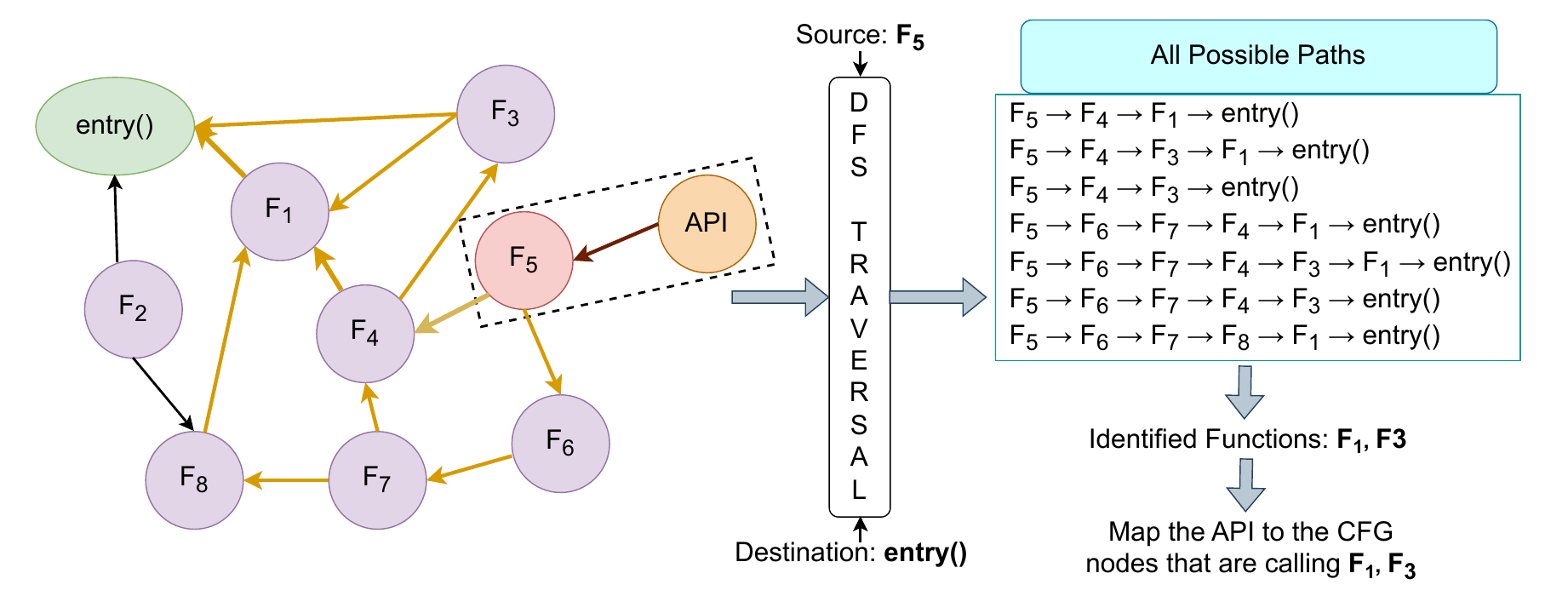}
\caption{Mapping of APIs to the CFG nodes.}
\label{mapper}
\end{figure}

We divide API mapping into two categories, namely (i) Direct references, where API calls are present within the CFG node, and (ii) Indirect references, where the actual API call occurs after a series of function calls initiated from the CFG node. Identifying direct references is usually simpler than identifying indirect ones, as they are contained within the CFG node itself. By retrieving the address of the API, we ascertain which CFG node it belongs to. However, identifying indirect references is challenging due to its recursive nature. For that, we begin by executing the command ``$ afl\sim[3-4]$" in the radare2 disassembled binary to identify all function calls within the binary. Next, we use the command ``$axt\ @$$<function\_call\_name>$" to explore the cross-references associated with each function call. This information allows us to create a directed graph that captures the function call names and their respective cross-references, as depicted in Figure \ref{mapper}. Subsequently, we execute the `$iij$` command to retrieve details about the import APIs and their associated Procedure Linkage Table (PLT) addresses within the binary. We then carry out a Depth First Search (DFS) traversal on the constructed graph to uncover all potential paths from the source to the destination. In this scenario, the source refers to the cross-reference function address associated with the API PLT address, and the destination is the `entry()` function. After obtaining the paths, we extract the function addresses along each route that are the immediate neighbors of the `entry()` function. Following compiling the function address list, we determine the CFG nodes that invoke these functions and map the API to the identified CFG nodes. This procedure is repeated for all the APIs present in the binary. As shown in Figure \ref{mapper}, API is the import API in the binary with $F_5$ function as the PLT cross-reference points. We apply DFS traversal, taking the $F_5$ as the source and \textit{entry()} as the destination to find all the possible paths. We identify seven paths available from $F_5$ to the \textit{entry()} function. Out of all the identified paths, the functions $F_1$ and $F_3$ are the immediate function calls before the \textit{entry()}. Following that, we map the API
to the nodes in the CFG that are calling $F_1$ and $F_3$ functions. We exercise a similar procedure to map the malicious strings to the nodes in the CFG.\\
\subsection{ML Classifier}
Neural networks capture intricate relationships between features, solve complex problems, and mimic human-like intelligence. They have a wide range of applications, including image recognition, fraud detection, medical diagnosis, and sentiment classification. Therefore, we use Deep Neural Network (DNN) to train and test the generated advanced features in the proposed work. We use `\textit{StandardScaler}' to normalize the features,`\textit{Adam}' as an optimizer, and `\textit{binary cross entropy}' as the loss function. We use the DNN model with thirteen layers (DNN13), where all the hidden layers are fully connected. We do batch normalization and node dropout before passing them to the next layer to speed up the training time and avoid overfitting \cite{overfast}, respectively. The number of output neurons in each layer of DNN13, starting from layer-1 to layer-13, is 5608, 5096, 4584, 4072, 3560, 3048, 2536, 2024, 1012, 512, 256, 128, and 1, respectively. We utilize ``\textit{Sigmoid}" as the activation function in the final layer as the chosen problem is a binary classification problem, whereas ``\textit{ReLU}" in the previous levels. The complete framework of the DNN is shown in Algorithm \ref{algorithm1}, and the descriptions of the notations used in Algorithm \ref{algorithm1} are mentioned in Table \ref{notationsss}.
\begin{algorithm}[]
\scriptsize
\caption{Neural Network Training Algorithm}\label{algorithm1}
\KwData{Random Waits \textit{W}, Random Biases \textit{b}, Input \textit{X}}
\KwResult{Trained Model}
\For{each training epoch}
{
    \For{each batch}
    {
        \For{each element in the batch}
        {
            \For{layer \textit{i} from \textit{1} to \textit{l}}
            {
                \For{neuron \textit{j} in layer \textit{i} ($j = 1$ to $n_i$)}
                {   
                    \If{i==1}
                    {
                        $Z_j^i = \sum_{k=1}^{s} W_{j,k}^i.X_k+b_j^i$\;
                        $a_j^{i} = ReLU(Z_j^i)$\;
                    }
                    \ElseIf{i==l}
                    {
                        $Z_j^i = \sum_{k=1}^{n_{i-1}} W_{j,k}^i.a_k^{i-1}+b_j^i$\;
                        $a_j^{i} = Sigmoid(Z_j^i)$\;
                    }
                    \Else
                    {
                        $Z_j^i = \sum_{k=1}^{n_{i-1}} W_{j,k}^i.a_k^{i-1}+b_j^i$\;
                        $a_j^{i} = ReLU(Z_j^i)$\;
                    }
                    
                }
            } 
           
            $\underset{loss}{L} = -\frac{1}{n_l}\sum_{j=1}^{n_l}(y_j.\log(a_j^l)+(1-y_j).\log(1-a_j^l))$\;
            
            \For{layer \textit{i} from \textit{l} to \textit{1}}
            {
                \For{neuron \textit{j} in layer \textit{i} ($j = 1$ to $n_i$)}
                {
                    \If{i==l}
                    {
                        $\delta_j^i = (a_j^i-y_j).Sigmoid'(Z_j^i)$\;
                    }
                    \Else
                    {
                        \If{$Z_j^i>0$}
                        {
                            $\delta_j^i = (W_{j,:}^{i+1})^T.\delta^{i+1}.1$\;
                        }
                        \Else
                        {
                            $\delta_j^i = 0$\;
                        }
                        
                    }
                    $\triangle W_{j,:}^{i} = \delta_j^i.(a^{i-1})^T$\;
                    $Wsum_j^i = Wsum_j^i + \triangle W_{j,:}^{i}$\;
                    $\triangle b_{j}^{i} = \delta_j^i$\;
                    $bsum_j^i = bsum_j^i + \triangle b_{j}^{i}$\;
                }
            }
        }

        \For{layer \textit{i} from \textit{l} to \textit{1}}
            {
                \For{neuron \textit{j} in layer \textit{i} ($j = 1$ to $n_i$)}
                {
                    $W_{j,:}^{i} = W_{j,:}^{i} - \eta.\frac{Wsum_j^i}{m}$\;
                     $b_j^i = b_j^i - \eta.\frac{bsum_j^i}{m}$\;
                    \
                }
            }
    }
}
\end{algorithm}
\begin{table}[t]
\scriptsize
\centering
\caption{Neural Network Algorithm Notations}
\begin{tabular}{|l|l|l|l|}
\hline
\multicolumn{1}{|c|}{\textbf{Notation}} & \multicolumn{1}{c|}{\textbf{Description}}                                                                                                                 & \multicolumn{1}{c|}{\textbf{Notation}} & \multicolumn{1}{c|}{\textbf{Description}}                                                                                        \\ \hline
$l$, $m$                              & Number of layers, Batch size                                                                                                                           & $n_i$                                & Neurons in the $i^{th}$ layer                                                                                                 \\ \hline
$Z_j^i$                               & \begin{tabular}[c]{@{}l@{}}Weighted sum for $j^{th}$\\ neuron in $i^{th}$ layer\end{tabular}                                                           & $W_{j,k}^{i}$                        & \begin{tabular}[c]{@{}l@{}}Weight of edge $k$ of \\ neuron $j$ at layer $i$\end{tabular}                                      \\ \hline
$X_k$                                 & Element $k$ of input $X$                                                                                                                               & $s$                                  & Length of input $X$                                                                                                           \\ \hline
$b_j^i$                               & \begin{tabular}[c]{@{}l@{}}Bias of $j^{th}$ neuron \\ at layer $i$\end{tabular}                                                                        & $a_j^i$                              & \begin{tabular}[c]{@{}l@{}}Activation output of \\ neuron $j$ at layer $i$\end{tabular}                                       \\ \hline
$\underset{loss}{L}$                  & Binary cross-entropy loss                                                                                                                              & $y_j$                                & Actual output of neuron $j$                                                                                                   \\ \hline
$\delta_j^i$                          & \begin{tabular}[c]{@{}l@{}}Gradient at $j^{th}$ neuron\\ in layer $i$\end{tabular}                                                                     & $\triangle W_{j,:}^{i}$              & \begin{tabular}[c]{@{}l@{}}Weight update for $j^{th}$ \\ neuron at $i^{th}$ layer for \\ each element in a batch\end{tabular} \\ \hline
$Wsum_j^i$                            & \begin{tabular}[c]{@{}l@{}}Sum of weight updates for $j^{th}$\\ neuron for a batch - at the beginning \\ of each batch, it is set to zero\end{tabular} & $\triangle b_j^i$                    & \begin{tabular}[c]{@{}l@{}}Bias update for $j^{th}$ \\ neuron at $i^{th}$ layer \\ for each element in a batch\end{tabular}   \\ \hline
$bsum_j^i$                            & \begin{tabular}[c]{@{}l@{}}Sum of bias updates for $j^{th}$ \\ neuron for a batch - at the beginning \\ of each batch, it is set to zero\end{tabular}  & $\eta$                               & Learning rate                                                                                                                 \\ \hline
\end{tabular}
\label{notationsss}
\end{table}
\section{EXPERIMENTS AND RESULTS}
\label{section4}
\subsection{Experimental Setup and Evaluation Metrics}
We perform feature extraction on a Parallels VM running Ubuntu 20.04 on a MacBook Air M1. We conduct the experiments on a supercomputer and a Windows 11 Pro Machine. Table \ref{table5} shows the details of the setup and tools used. We use Python for scripting and recursive disassembler tools to extract the CFG as mentioned in Table \ref{table5}. We use JSON format to store the extracted API list and the string list of the binary. We use StringSifter \cite{ref10} for identifying the top strings with maliciousness scores. For normalizing the string-based features, we use FeatureHasher \cite{featurehasher}. We employ accuracy, precision, recall, Area Under ROC Curve (AUC), F1 score, FPR, and loss as evaluation metrics to identify the best ML model.
\subsection{Dataset Used}
\begin{figure}[t]
    \centering
    \includegraphics[width=\linewidth, trim = 3cm 1.95cm 0cm 0cm, clip]{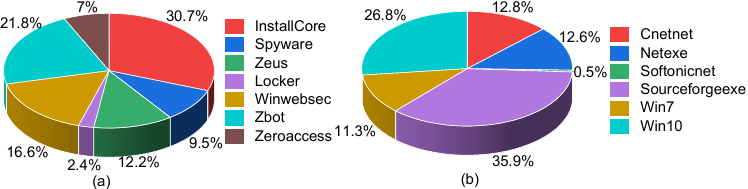}
    \caption{Dataset: (a) Malware (b) Benign.}
    \label{table6}
\end{figure}

To perform extensive analysis, we consider 9504 malware binaries \cite{ref7} and 11089 benign binaries from \cite{refdata} and freshly installed Windows 7 and Windows 10 machines. The malware composition includes spyware, ransomware, adware, trojans, and rogues. We verify the benign binaries with the VirusTotal Academic API key \cite{ref7}. Among the considered samples, we use 70\% samples for training and 30\% samples for testing the model. The details of the benign and malware binaries are shown in Figure \ref{table6}.
\begin{table}[!t]
\centering
\caption{Experimental Setup and Tool Used}
\scriptsize
\begin{tabular}{|l|l|}
\hline
\multicolumn{1}{|c|}{\textbf{Component}}                                                                                          & \multicolumn{1}{c|}{\textbf{Properties}}                                                                                                                                                                                        \\ \hline
\begin{tabular}[c]{@{}l@{}}Feature Extraction\end{tabular} & \begin{tabular}[c]{@{}l@{}}\textbf{Operating System:} Ubuntu 20.04 (guest machine), \\ \textbf{CPU details:} 2vCPUs of the VM and \\ 8 core CPU of the host machine\\ \textbf{Memory:} 2 GB RAM of the VM and \\ 16GB RAM of the host machine\end{tabular} \\ \hline
\begin{tabular}[c]{@{}l@{}}Model Evaluation\end{tabular} & \begin{tabular}[c]{@{}l@{}}\textbf{Operating System:} Ubuntu 20.04 LTS \\ \textbf{CPU details:} AMD EPYC 7763, 64-core processor,\\10 CPU cores, 
64 cores per socket, 128 threads,\\1.64 GHz base clock, 
512 KB cache per core\\ \textbf{Memory:} 200 GB RAM\\ \textbf{GPU:} NVIDIA A100 80GB PCIe\end{tabular} \\ \hline
\begin{tabular}[c]{@{}l@{}}Feature Importance\\ Calculation\end{tabular} & \begin{tabular}[c]{@{}l@{}}\textbf{Operating System:} Windows 11 Pro, \\ \textbf{CPU details:} 13th Gen Intel(R) Core(TM)\\i7-13700 2.10GHz.\\ \textbf{Memory:} 64 GB\\ \textbf{GPU:} NVIDIA T400 4GB\end{tabular} \\ \hline
Feature Extraction Module                                                                                                         & \textbf{Tools:} LIEF, Radare2, r2pipe, StringSifter                                                                                                                                                                              \\ \hline
Machine Learning Module                                                                                                           & \begin{tabular}[c]{@{}l@{}}\textbf{Tools:}  Scikit-learn, Keras, Tensorflow, Graphviz, \\ TensorBoard\end{tabular}                                                                                                                       \\ \hline
\end{tabular}
\label{table5}
\end{table}
\begin{figure}
    \centering
    \includegraphics[width=\linewidth, trim = 0cm 0cm 0cm 0cm, clip]{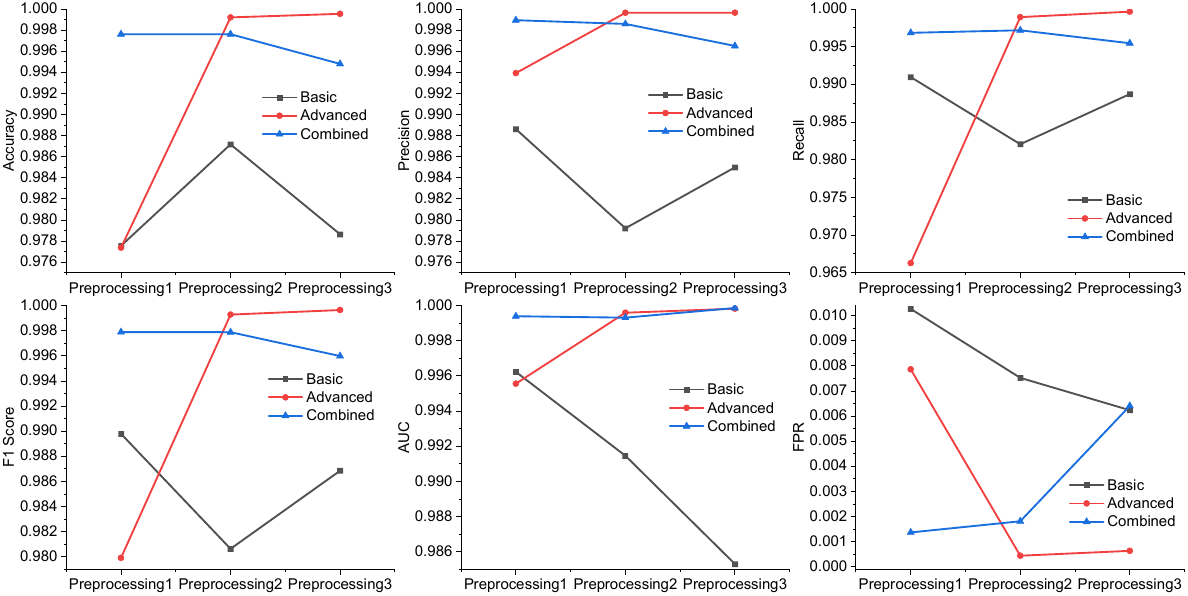}
    \caption{Experimental Comparison between Advanced, Basic, Advanced+Basic Features with Three Different Pre-processings.}
    \label{Figure05}
\end{figure}
\subsection{Results and Discussion}
We employ three different ways of preprocessing on the concatenated feature sets of advanced and basic features to test the robustness and the importance of advanced features over basic features. First, we remove all the duplicate rows from the feature set. We name this step as \textit{Duplicate Removal}. Second, we perform correlation-based feature selection besides removing the duplicated rows from the feature set, named this step as \textit{Correlated Features Filtering}. Finally, we remove the rows with the most frequently occurring advanced feature entry from the feature set to mitigate bias introduced by overrepresented patterns and to improve model generalization, called \textit{Dummy Removal}. 

We compare the advanced, basic, and advanced+basic features based on the chosen model evaluation metrics to understand the importance and quality of the advanced feature set. To do so, we implement and evaluate the DNN13 model on all the preprocessing steps. As shown in Figure \ref{Figure05}, the advanced features produce better results compared to basic features and combined features in terms of all the evaluation metrics by producing more than 99\% accuracy, precision, recall, F1-score, AUC, and 0.064\% FPR.

\begin{figure}[t]
    \centering
    \begin{subfigure}[b]{0.325\textwidth}
        \centering
        \includegraphics[width=\textwidth, trim = 0cm 0cm 0cm 0cm, clip]{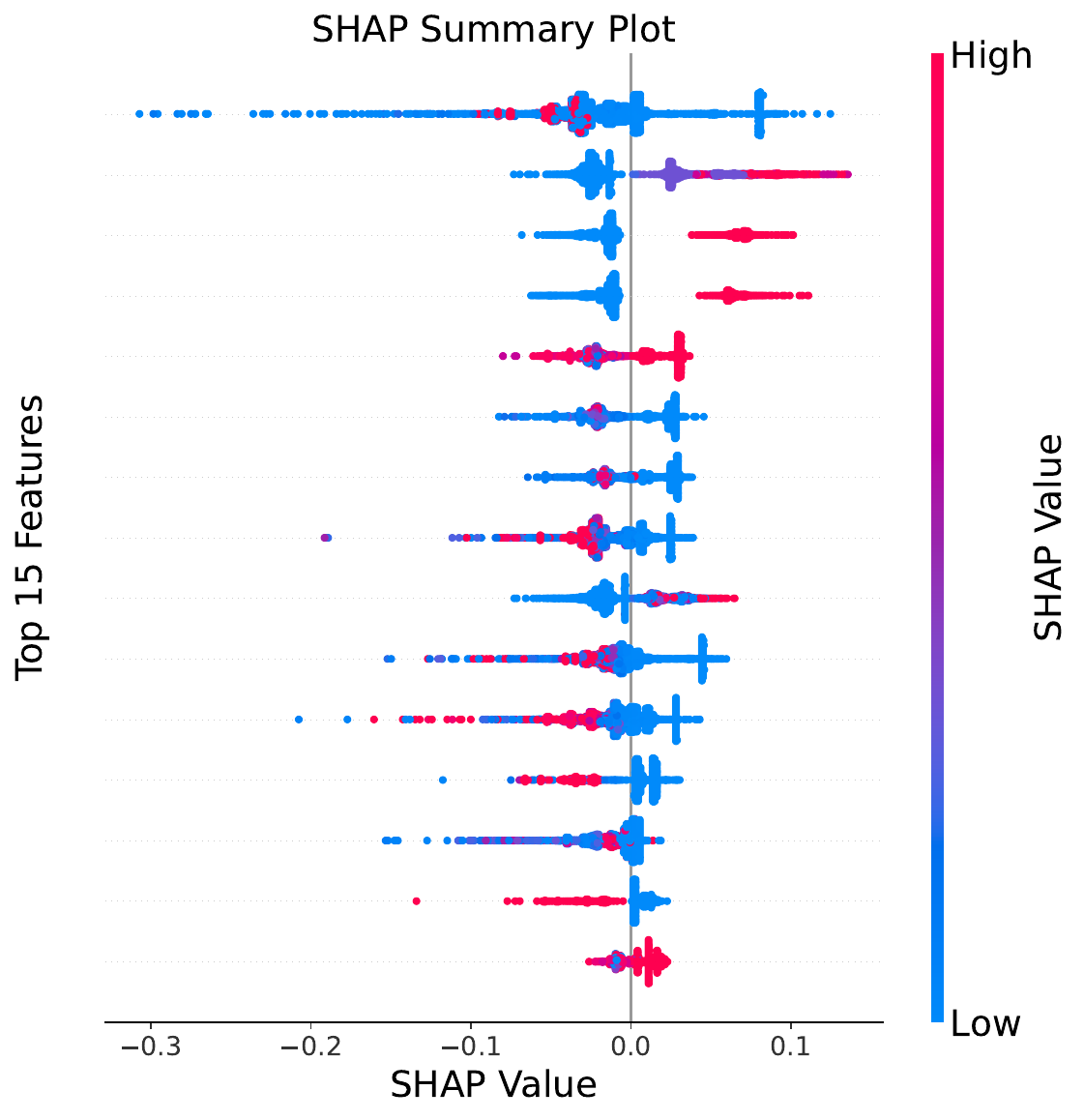}
        \caption{Advanced}
        \label{}
    \end{subfigure}
    \hfill
    \begin{subfigure}[b]{0.325\textwidth}
        \centering
        \includegraphics[width=\textwidth, trim = 0cm 0cm 0cm 0cm, clip]{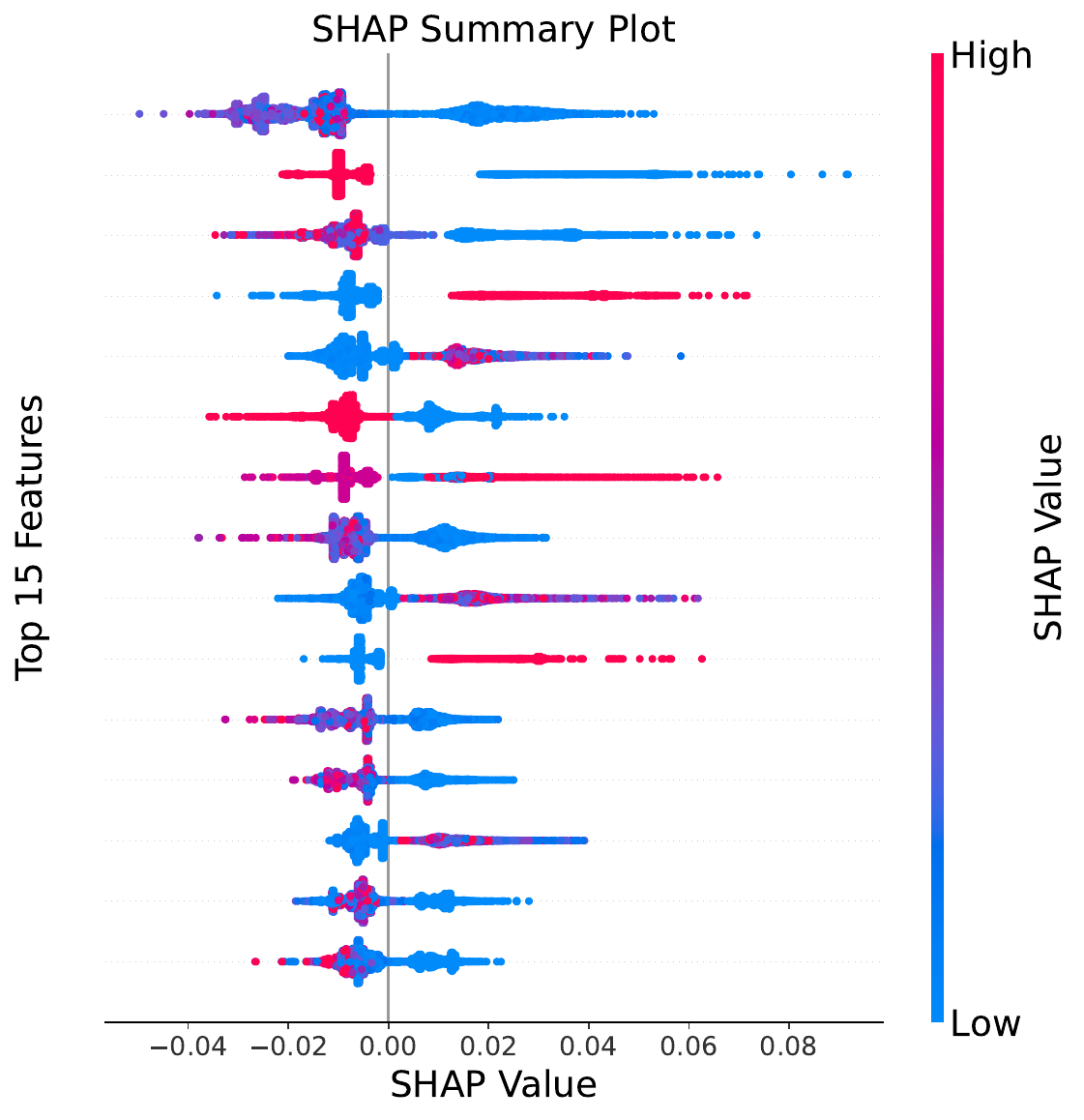}
        \caption{Basic}
        \label{}
    \end{subfigure}
    \hfill  
    \begin{subfigure}[b]{0.325\textwidth}
        \centering
        \includegraphics[width=\textwidth, trim = 0cm 0cm 0cm 0cm, clip]{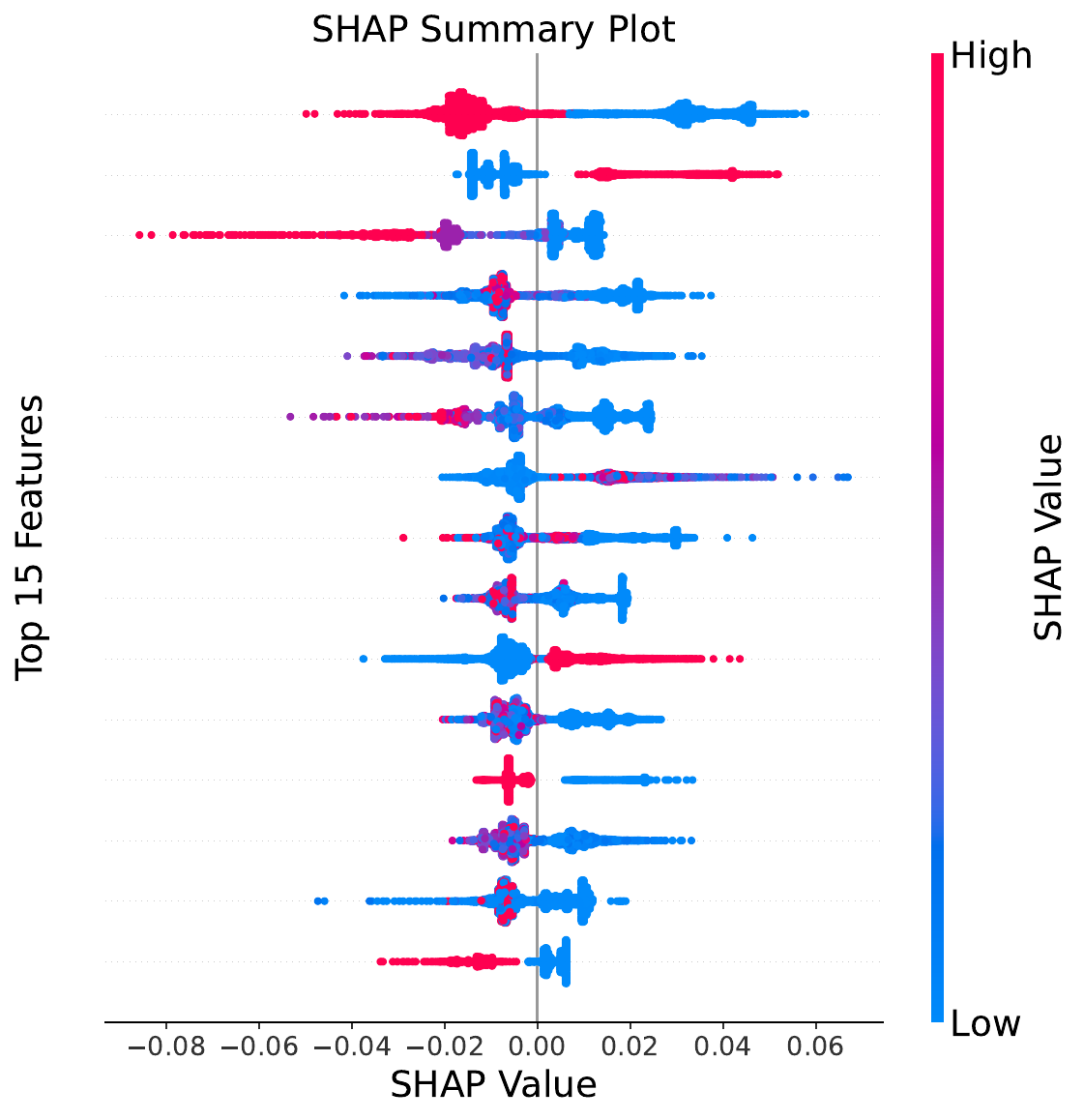}
        \caption{Combined}
        \label{}
    \end{subfigure}
    \hfill
    \begin{subfigure}[b]{0.325\textwidth}
        \centering
        \includegraphics[width=\textwidth, trim = 0cm 0cm 0cm 0cm, clip]{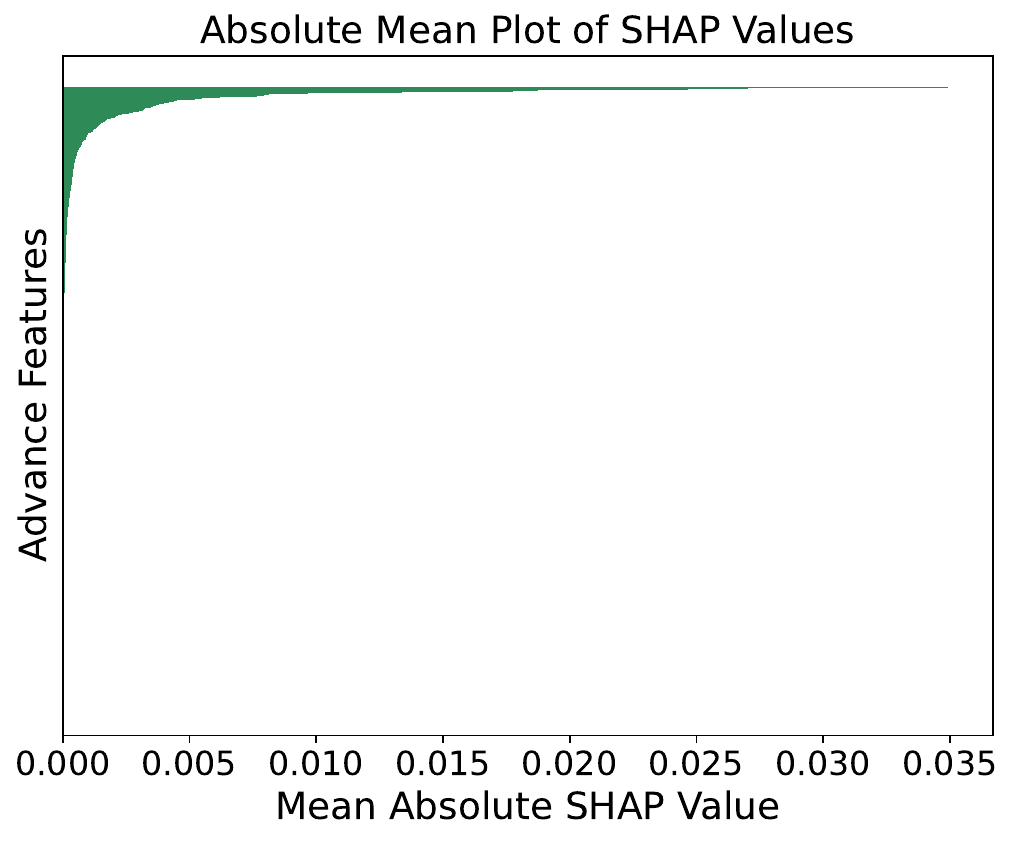}
        \caption{Advanced}
        \label{}
    \end{subfigure}
    \hfill
    \begin{subfigure}[b]{0.325\textwidth}
        \centering
        \includegraphics[width=\textwidth, trim = 0cm 0cm 0cm 0cm, clip]{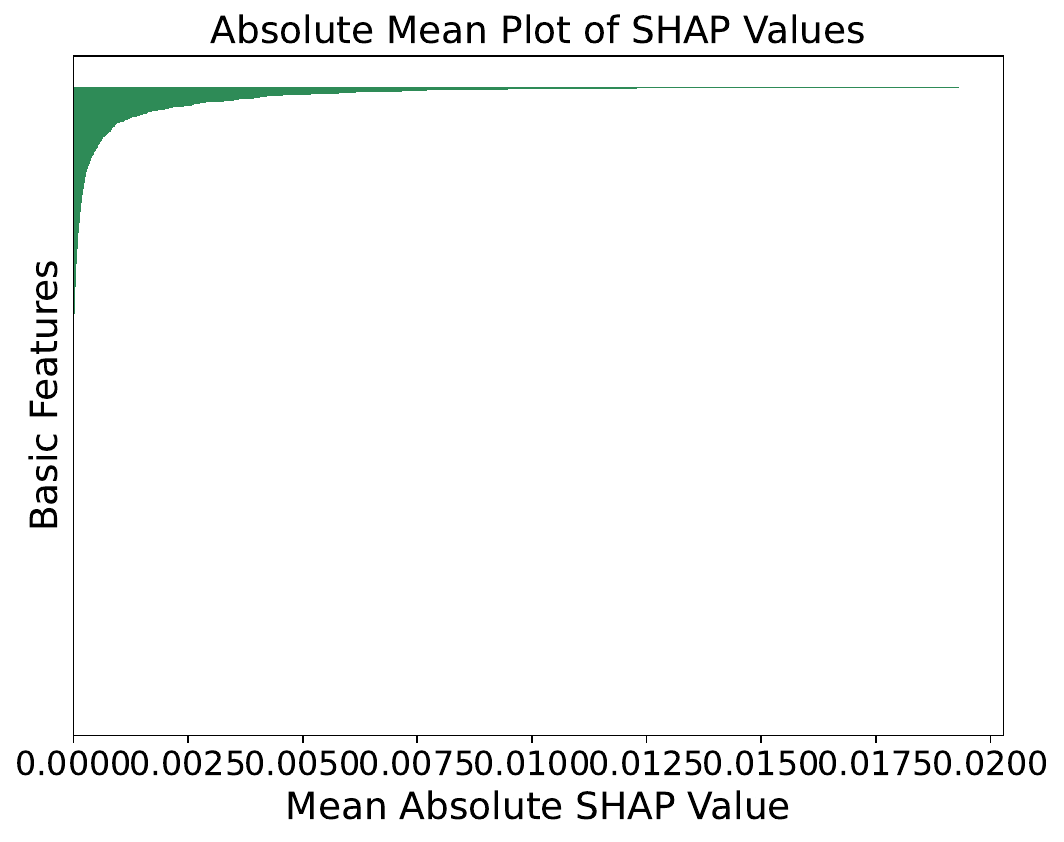}
        \caption{Basic}
        \label{}
    \end{subfigure}
    \hfill   
    \begin{subfigure}[b]{0.325\textwidth}
        \centering
        \includegraphics[width=\textwidth, trim = 0cm 0cm 0cm 0cm, clip]{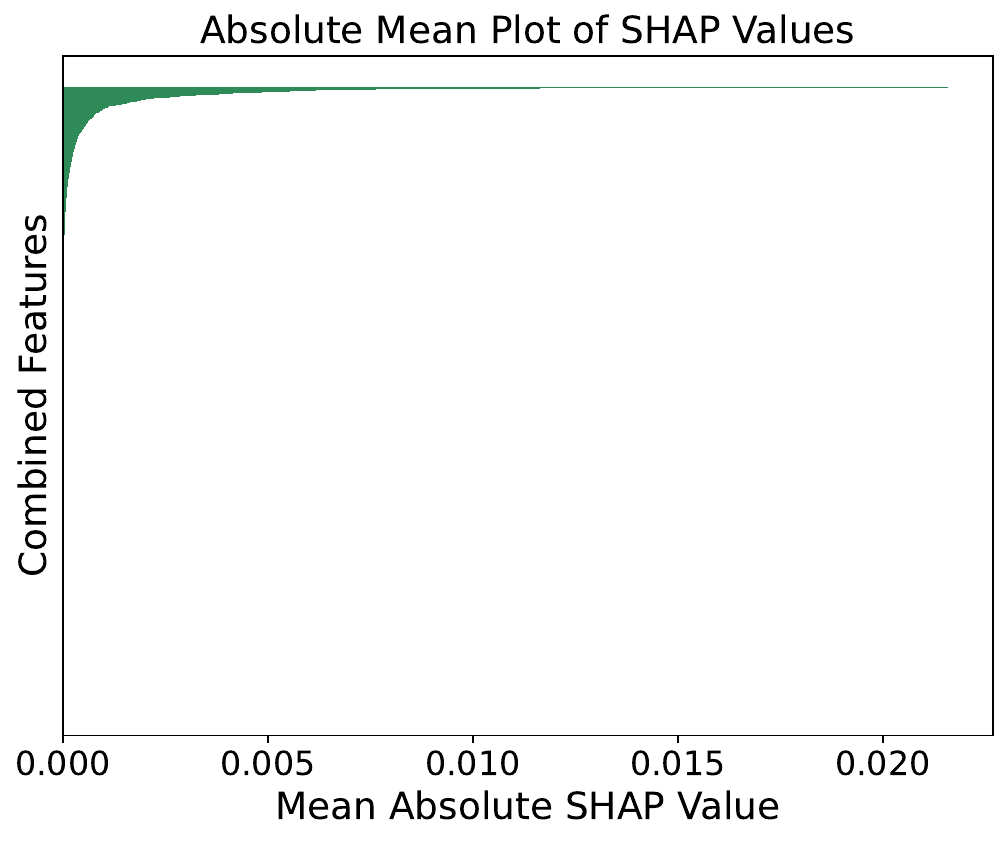}
        \caption{Combined}
        \label{}
    \end{subfigure}
    \caption{Beeswarm and Absolute Mean Plots on Preprocessing-1 Features.}
    \label{shaply_p1}
\end{figure}

\begin{figure}[t]
    \centering
    \begin{subfigure}[b]{0.325\textwidth}
        \centering
        \includegraphics[width=\textwidth, trim = 0cm 0cm 0cm 0cm, clip]{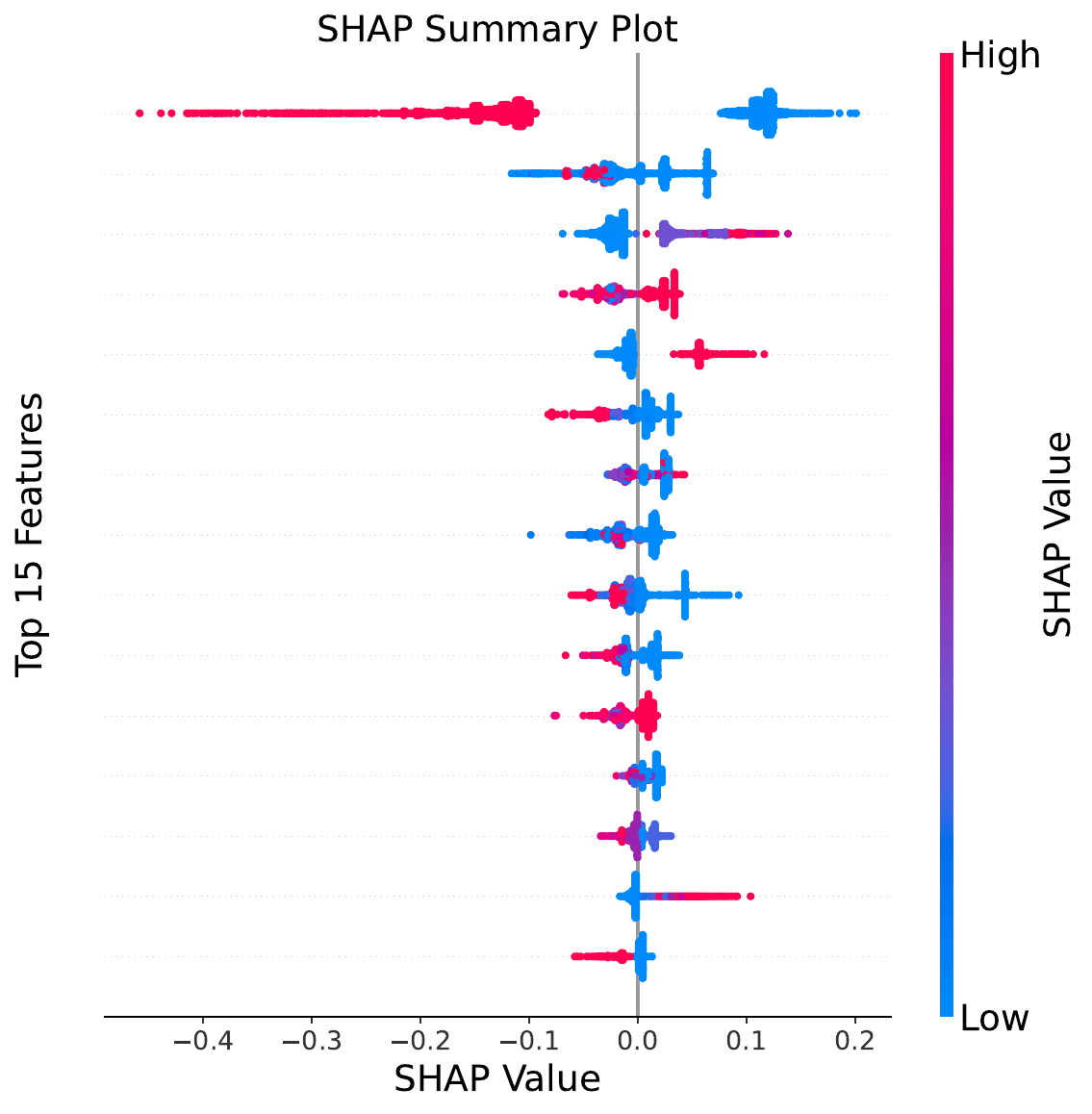}
        \caption{Advanced}
        \label{}
    \end{subfigure}
    \hfill
    \begin{subfigure}[b]{0.325\textwidth}
        \centering
        \includegraphics[width=\textwidth, trim = 0cm 0cm 0cm 0cm, clip]{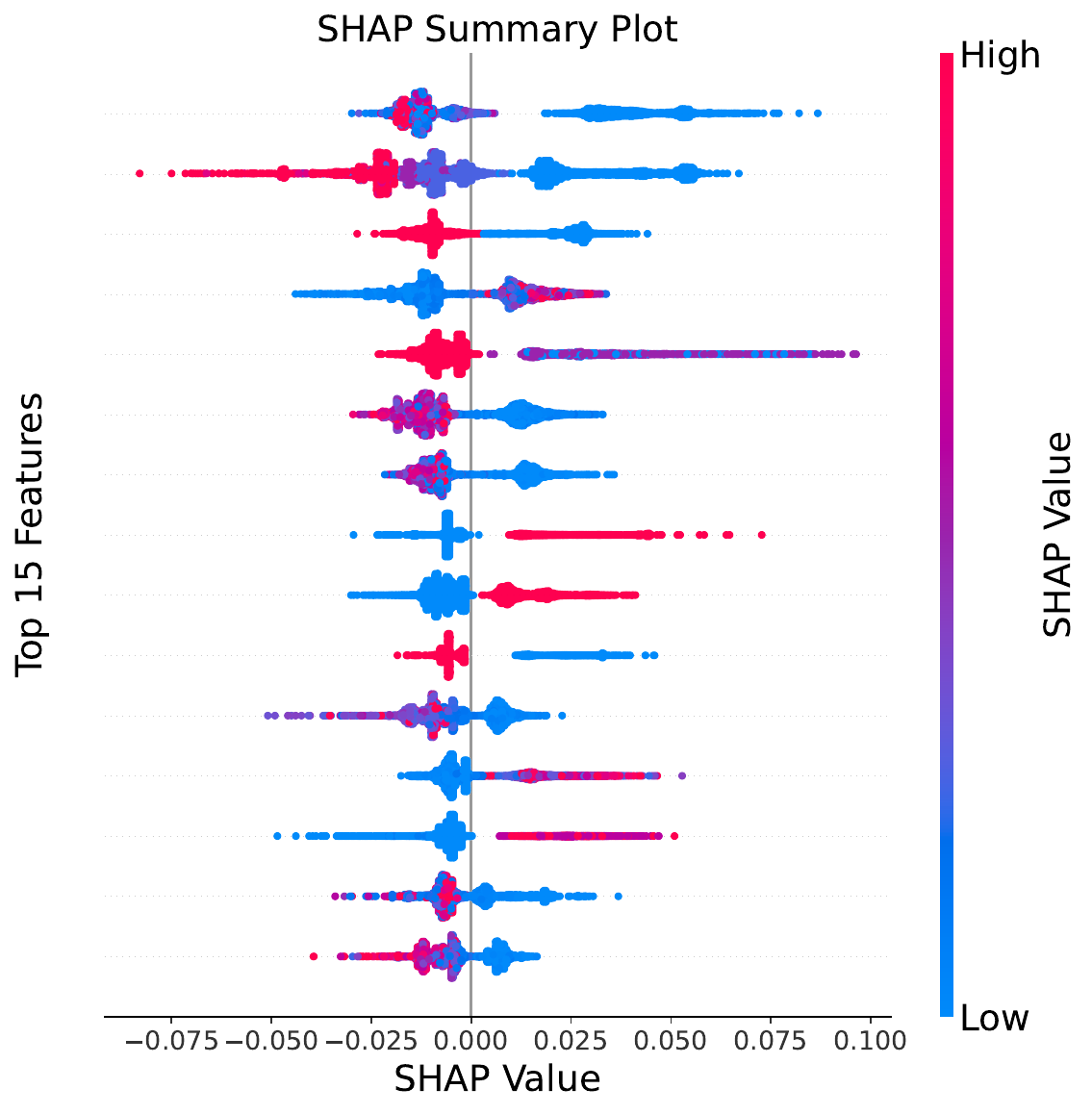}
        \caption{Basic}
        \label{}
    \end{subfigure}
    \hfill  
    \begin{subfigure}[b]{0.325\textwidth}
        \centering
        \includegraphics[width=\textwidth, trim = 0cm 0cm 0cm 0cm, clip]{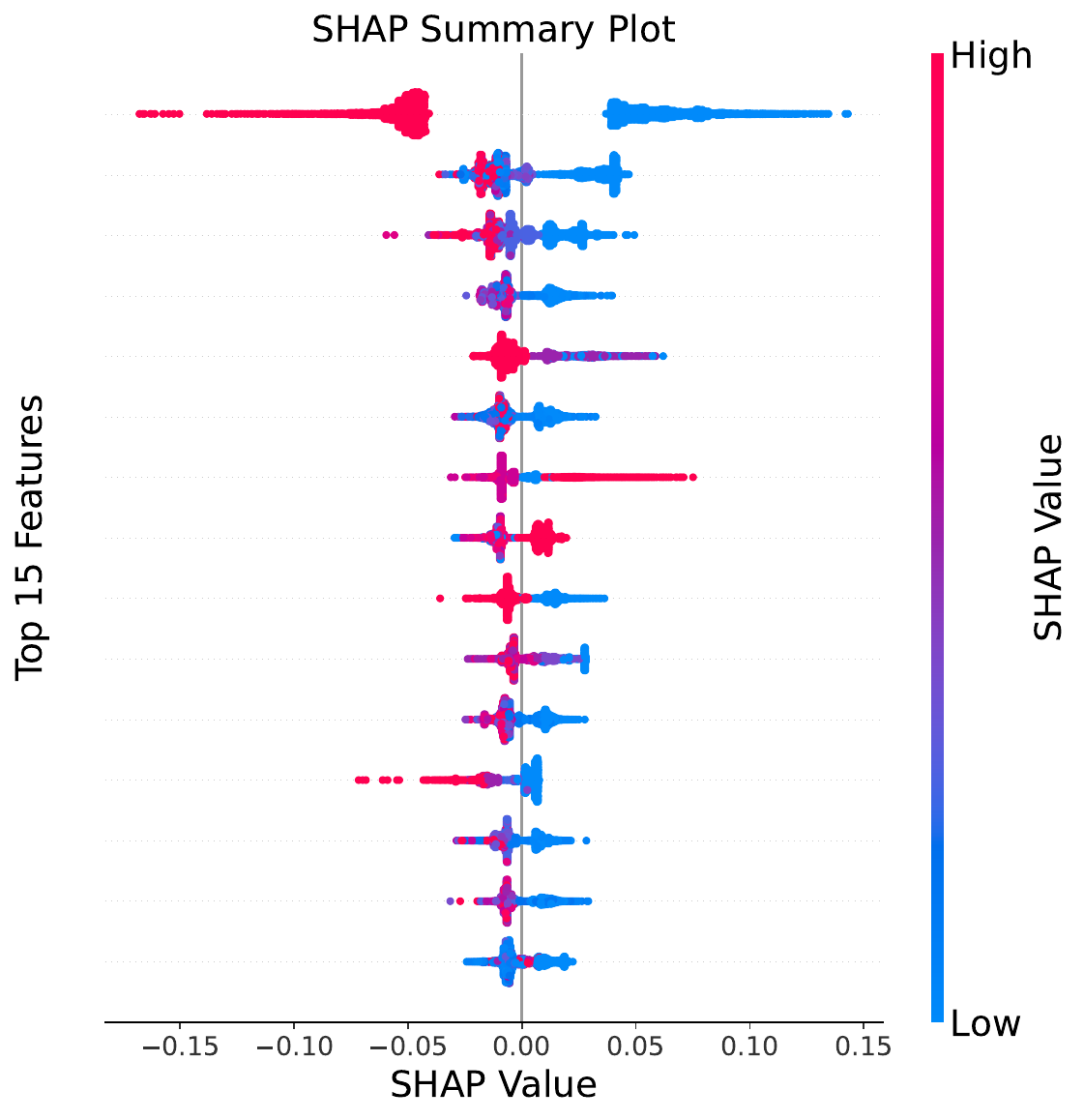}
        \caption{Combined}
        \label{}
    \end{subfigure}
    \hfill
    \begin{subfigure}[b]{0.325\textwidth}
        \centering
        \includegraphics[width=\textwidth, trim = 0cm 0cm 0cm 0cm, clip]{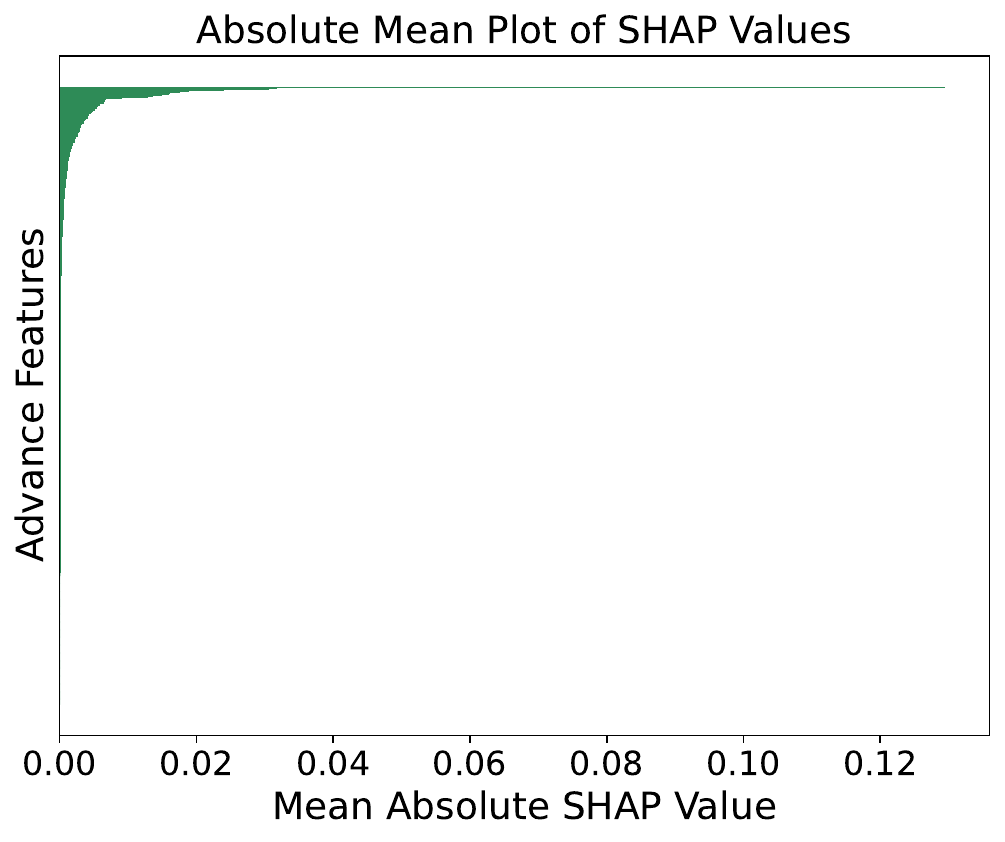}
        \caption{Advanced}
        \label{}
    \end{subfigure}
    \hfill
    \begin{subfigure}[b]{0.325\textwidth}
        \centering
        \includegraphics[width=\textwidth, trim = 0cm 0cm 0cm 0cm, clip]{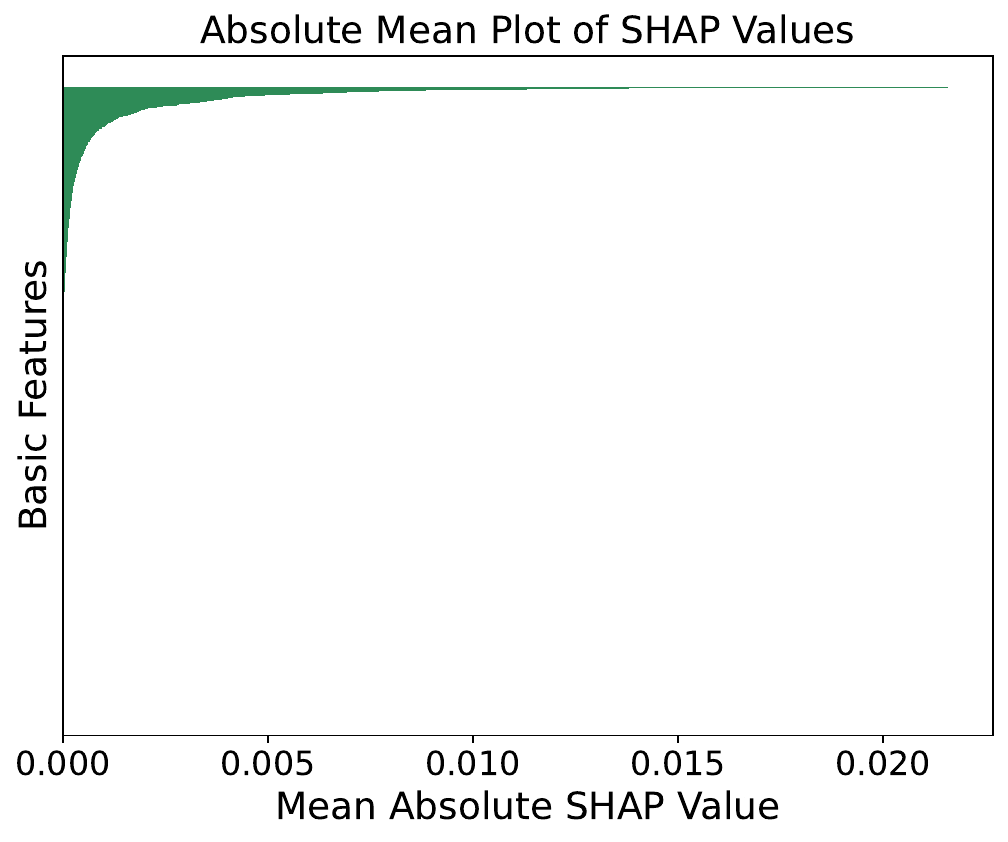}
        \caption{Basic}
        \label{}
    \end{subfigure}
    \hfill   
    \begin{subfigure}[b]{0.325\textwidth}
        \centering
        \includegraphics[width=\textwidth, trim = 0cm 0cm 0cm 0cm, clip]{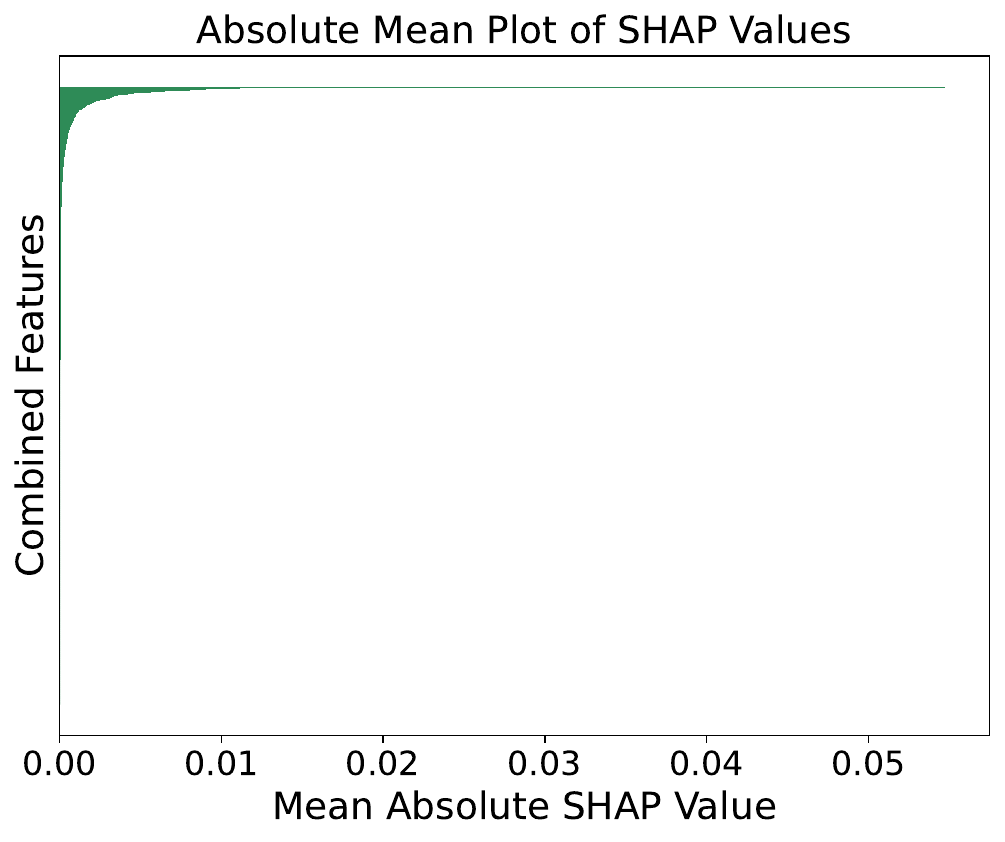}
        \caption{Combined}
        \label{}
    \end{subfigure}
    \caption{Beeswarm and Absolute Mean Plots on Preprocessing-2 Features.}
    \label{shaply_p2}
\end{figure}

\begin{figure}[t]
    \centering
    \begin{subfigure}[b]{0.325\textwidth}
        \centering
        \includegraphics[width=\textwidth, trim = 0cm 0cm 0cm 0cm, clip]{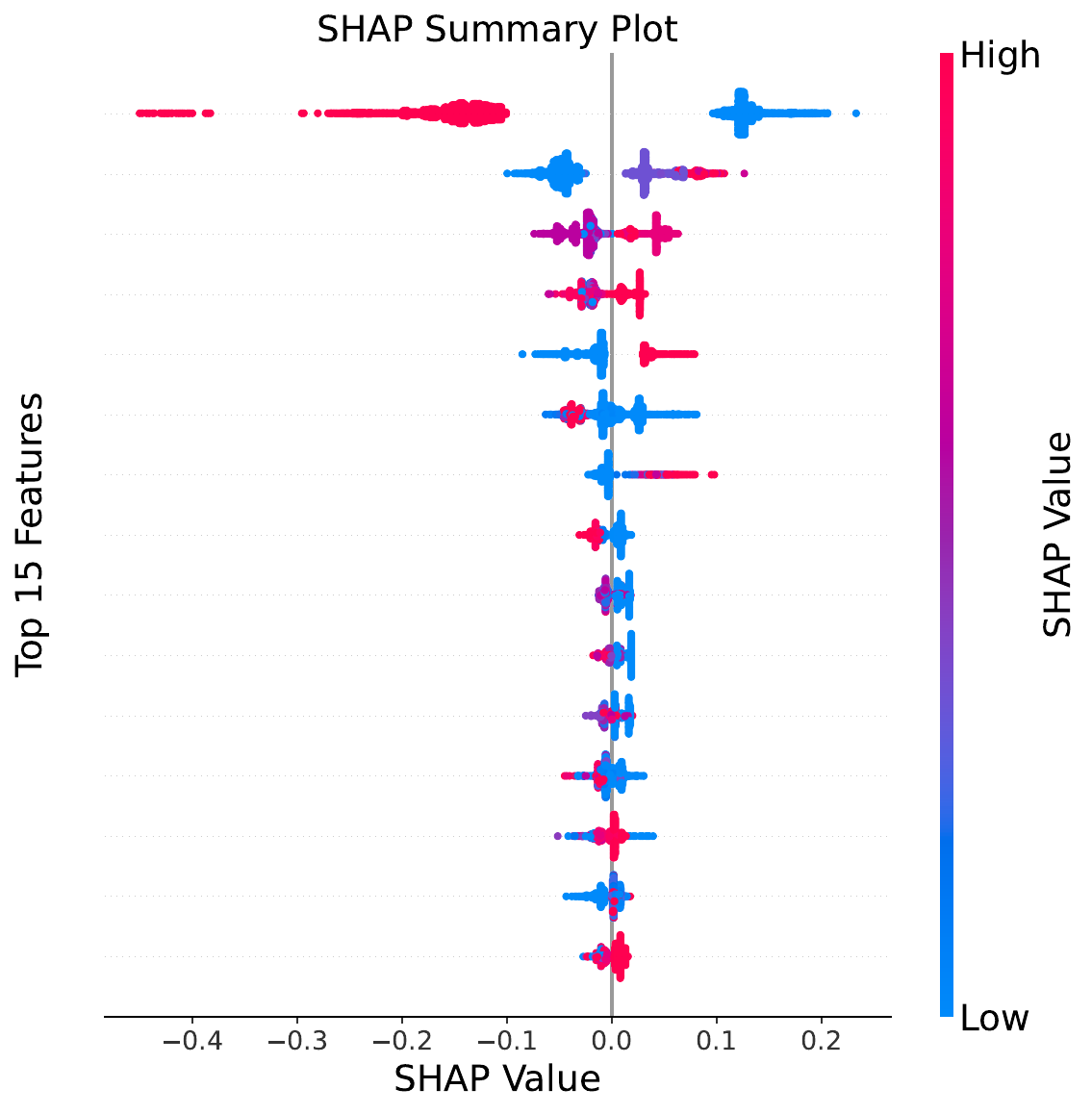}
        \caption{Advanced}
        \label{}
    \end{subfigure}
    \hfill
    \begin{subfigure}[b]{0.325\textwidth}
        \centering
        \includegraphics[width=\textwidth, trim = 0cm 0cm 0cm 0cm, clip]{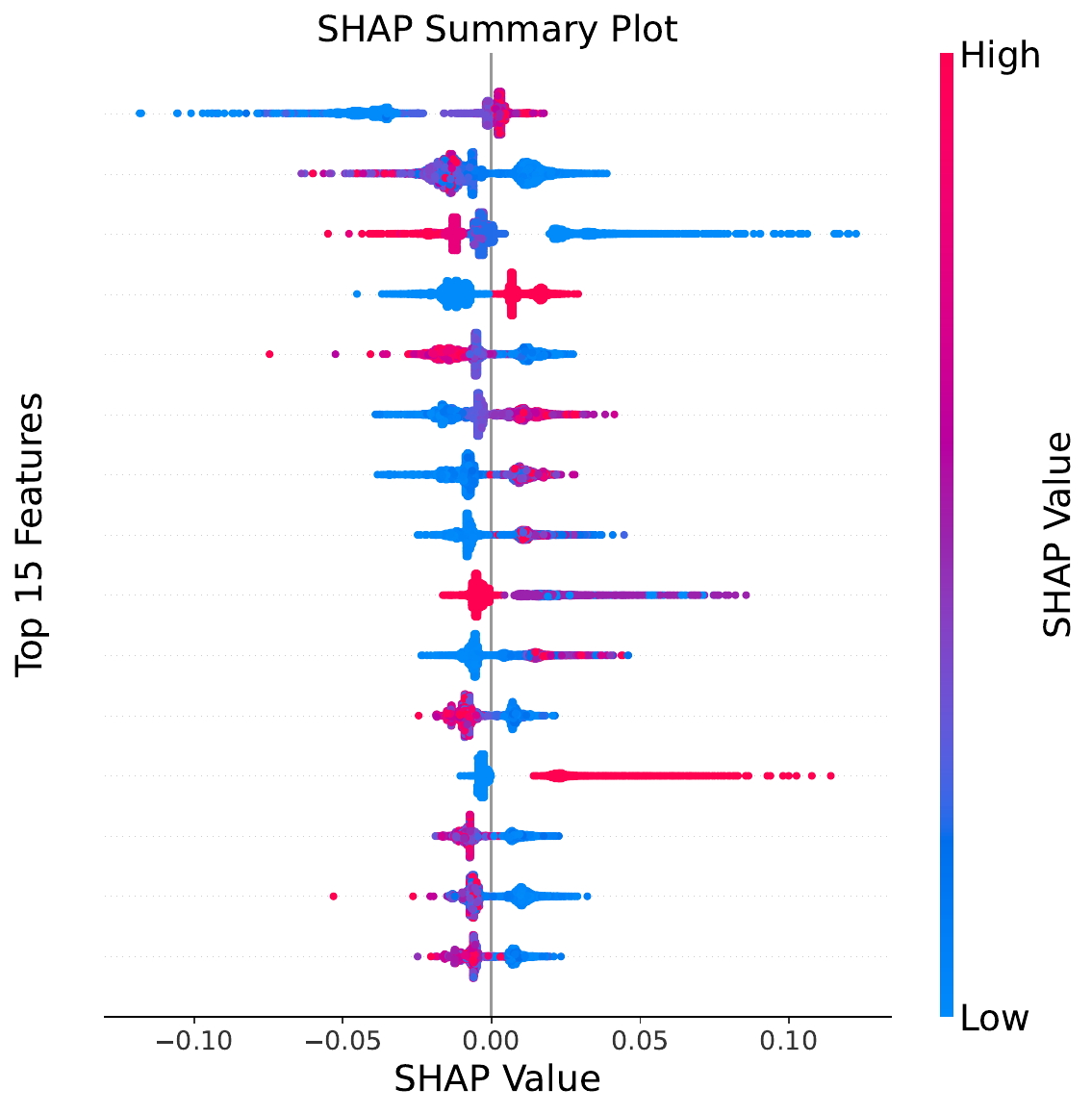}
        \caption{Basic}
        \label{}
    \end{subfigure}
    \hfill  
    \begin{subfigure}[b]{0.325\textwidth}
        \centering
        \includegraphics[width=\textwidth, trim = 0cm 0cm 0cm 0cm, clip]{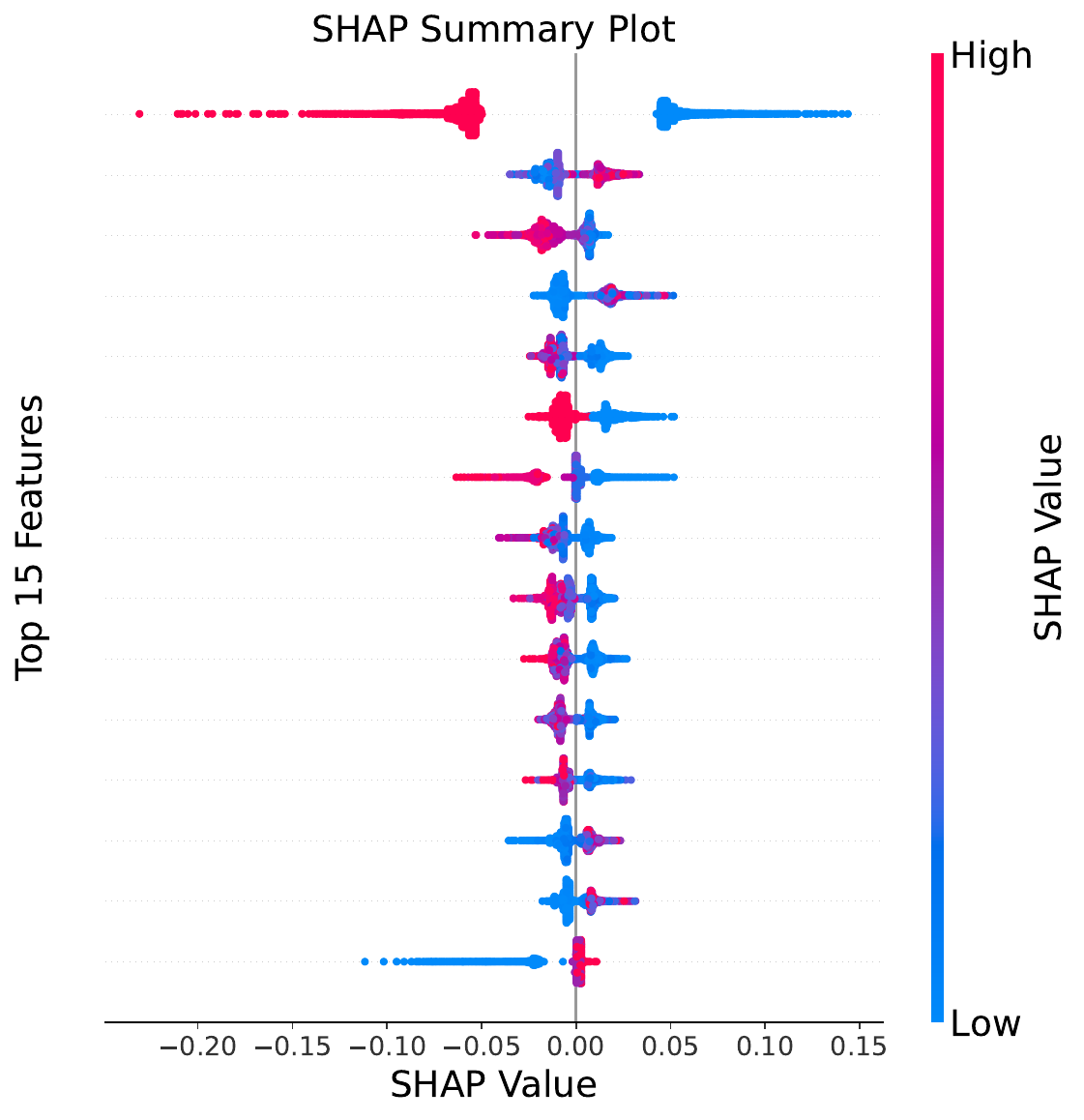}
        \caption{Combined}
        \label{}
    \end{subfigure}
    \hfill
    \begin{subfigure}[b]{0.325\textwidth}
        \centering
        \includegraphics[width=\textwidth, trim = 0cm 0cm 0cm 0cm, clip]{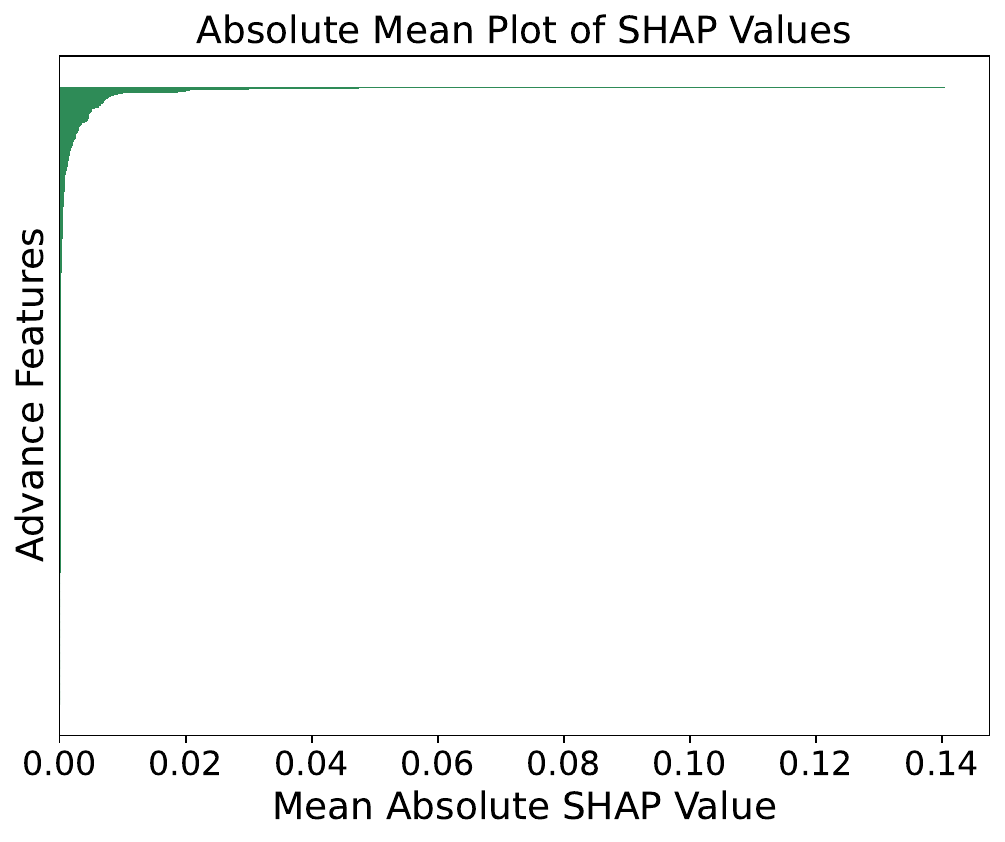}
        \caption{Advanced}
        \label{}
    \end{subfigure}
    \hfill
    \begin{subfigure}[b]{0.325\textwidth}
        \centering
        \includegraphics[width=\textwidth, trim = 0cm 0cm 0cm 0cm, clip]{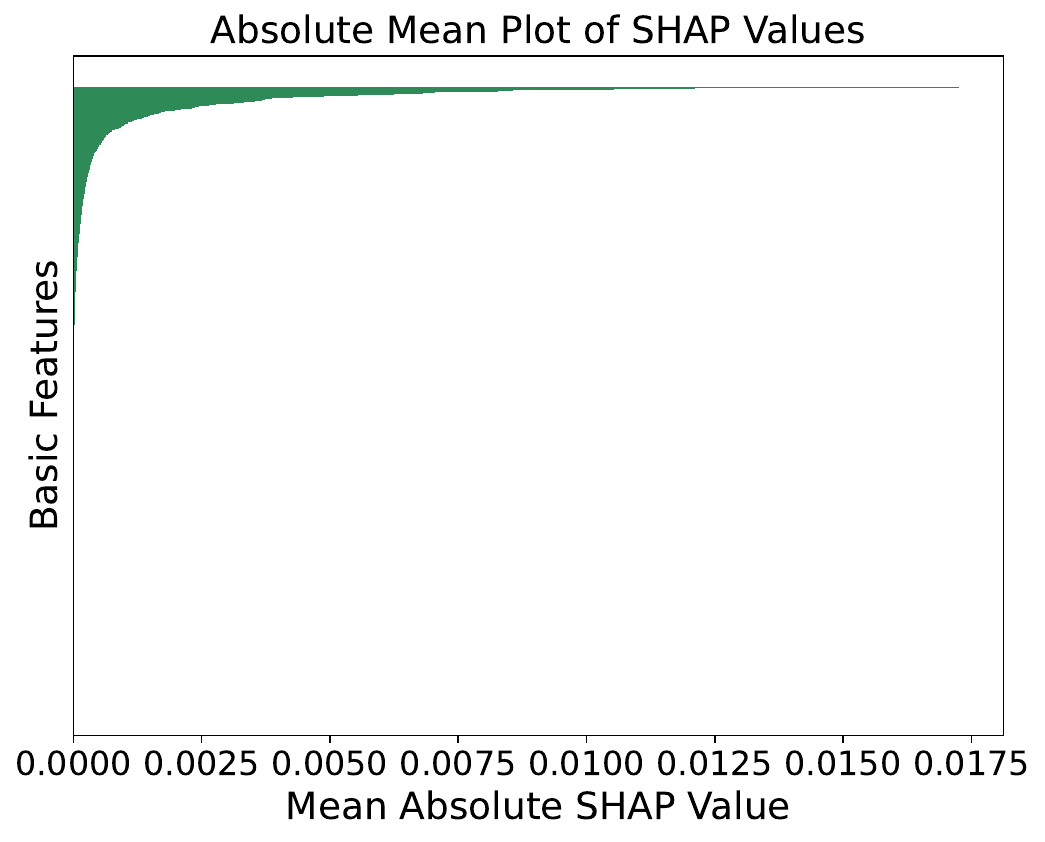}
        \caption{Basic}
        \label{}
    \end{subfigure}
    \hfill   
    \begin{subfigure}[b]{0.325\textwidth}
        \centering
        \includegraphics[width=\textwidth, trim = 0cm 0cm 0cm 0cm, clip]{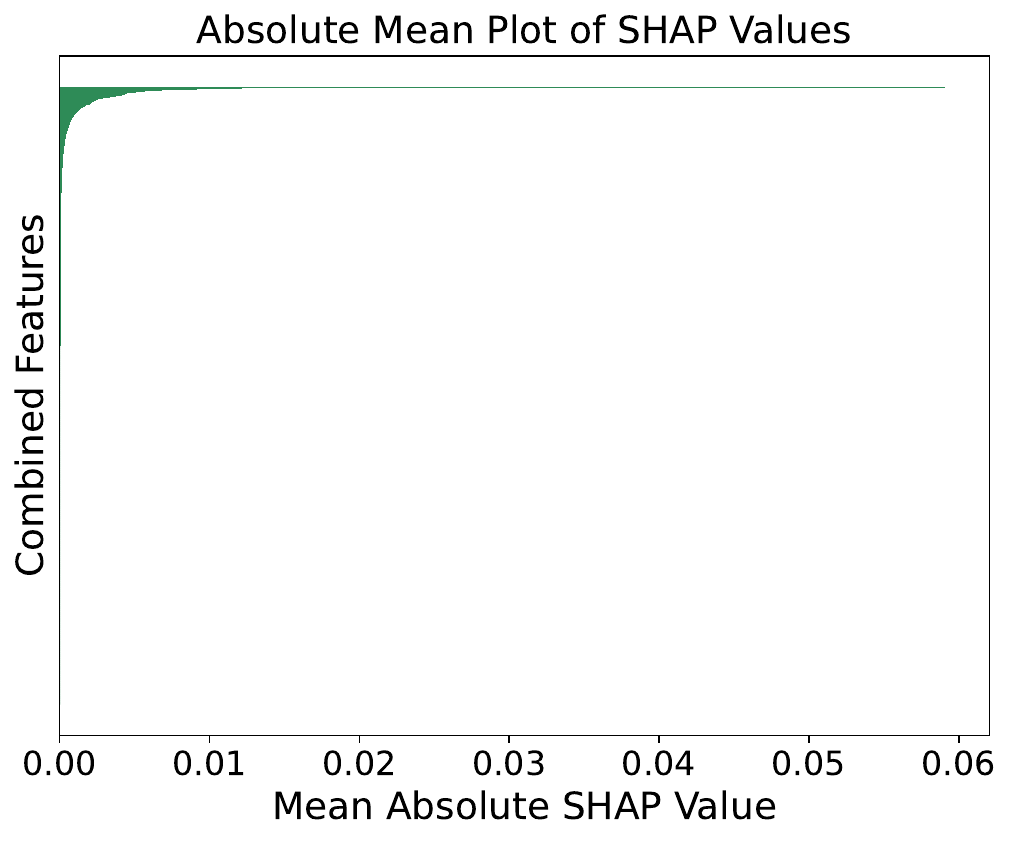}
        \caption{Combined}
        \label{}
    \end{subfigure}
    \caption{Beeswarm and Absolute Mean Plots on Preprocessing-3 Features.}
    \label{shaply_p3}
\end{figure}

\textit{Measuring Feature Importance:} SHapley Additive exPlanations (SHAP) calculates the contribution of each feature in the overall prediction. SHAP measures the influence of each feature on a prediction by evaluating all possible combinations of features. It assigns a value to each feature based on how much it contributes to increasing or decreasing the predicted outcome when it is included in the model compared to when it is not. One of the key advantages of SHAP is that it ensures consistency and fairness in feature attribution by reflecting the contribution of features based on the model's behavior. Therefore, we apply SHAP on advanced, basic, and combined features to measure which one contributes the most. As shown in Figure \ref{shaply_p1}, Figure \ref{shaply_p2}, and Figure \ref{shaply_p3}, we plot Beeswarm and Absolute Mean plots for advanced, basic, and combined features on preprocessing-1, preprocessing-2, and preprocessing-3. The Absolute Mean plots for three preprocessing show that advanced features contribute 5.68\%, 1.15\%, and 8.13\% more than basic features and 3.60\%, 3.48\%, and 9.88\% more than combined features. The Beeswarm plots for three preprocessing show that advanced features contribute 0\%, 1.47\%, and 1.44\% more than basic features and 0.05\%, 2.51\%, and 2.42\% more than combined features. Further, attesting the importance of extracting the features from potential regions of maliciousness and completely preprocessed CFG.  

\begin{figure}[t]
\centering
\includegraphics[width=\textwidth, trim = 0cm 0.5cm 0cm 0cm, clip]{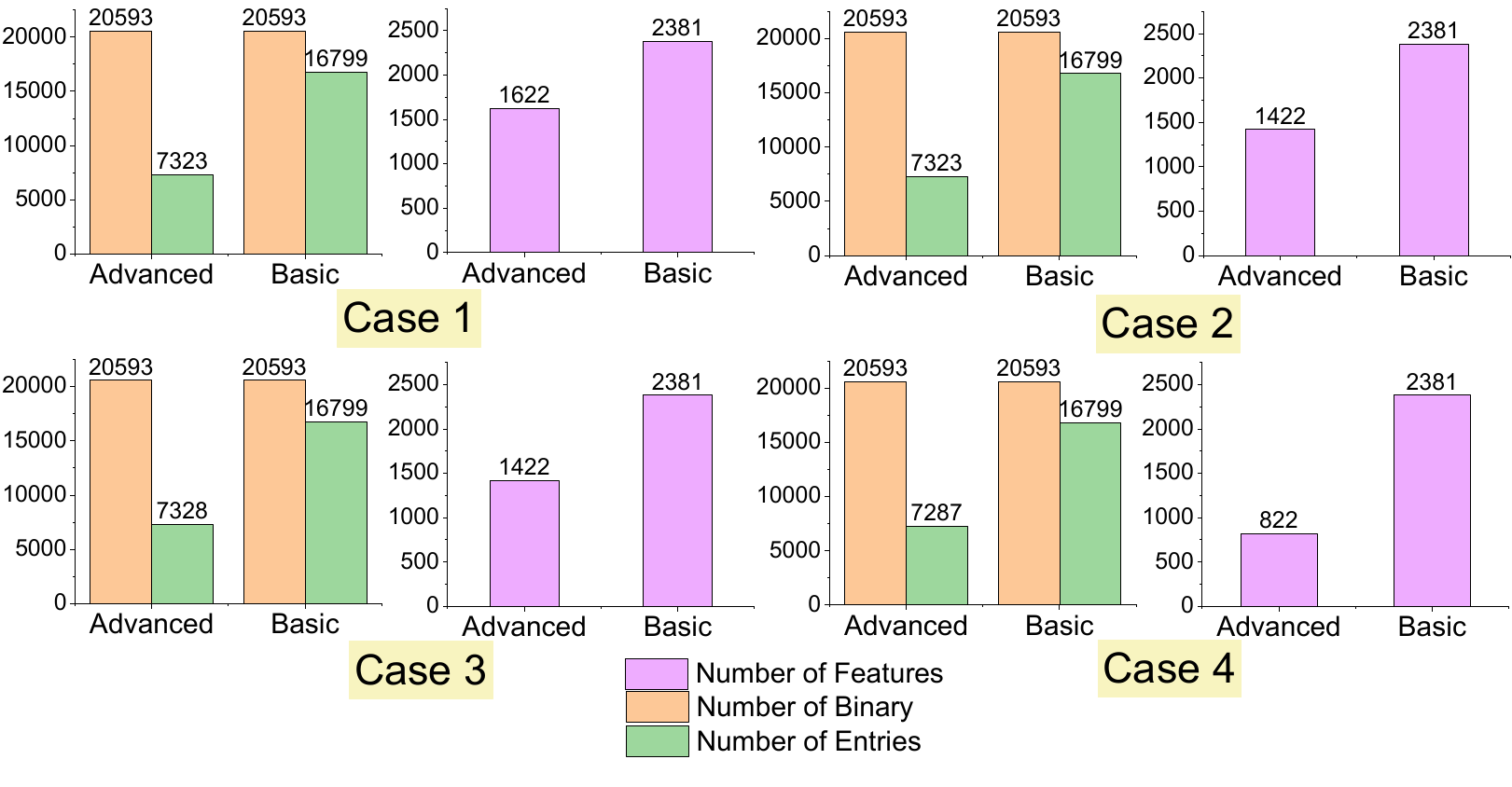}
\caption{Comparison of PotentRegion4MalDetect and a model that focuses on the entire binary based on the number of features and entries.}
\label{features_entries}
\end{figure}

\textit{Comparison of PotentRegion4MalDetect with a model that focuses on the entire binary:} We evaluate the PotentRegion4MalDetect and the model that focuses on the entire binary by comparing the number of entries for binaries with the same functionality and the total number of features each generates. As shown in Figure \ref{features_entries}, the proposed model requires lesser entries than the model focusing on the entire binary. It emphasizes that the proposed model creates the same entries for the binaries of the same behavior, as it focuses on regions of potential maliciousness and completely preprocessed CFG for feature extraction. However, the model that extracts features from the entire binary treats binaries as different and creates separate entries despite the binaries having the same core functionalities. Further, the proposed model focuses on a specific number of robust features rather than more shallow features that can be evaded by advanced malware. These two properties of the proposed model over the models that focus on the entire binary improve the data handling, reduce memory overhead, allow for faster computation, and lower storage requirements. This answers the research question (R5).
\begin{figure}[!]
	\centering
	\includegraphics[width=\textwidth, trim = 0cm 0cm 0cm 0cm, clip]{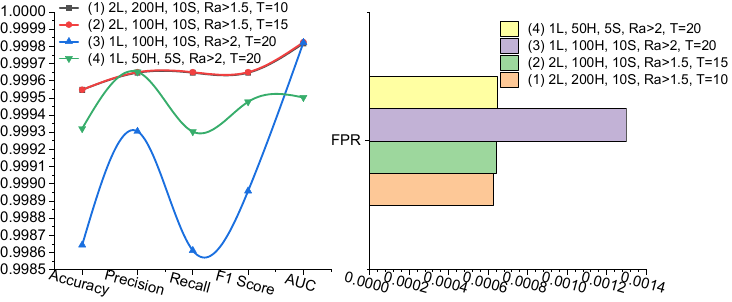}
	\caption{Experiments with Different Hyperparameters: Level (L), Feature Hasher (H), Subgraphs (S), Ratio (Ra), Trigrams (T), Accuracy (A), Precision (P), Recall (R), Area Under ROC Curve (AUC).}
	\label{table12}
\end{figure}

\textit{Experiments with Different Hyperparameters:} In this study, we conduct four distinct experiments, each featuring different hyperparameters, including the number of in-flow levels, the number of out-flow levels, the types of Feature Hashers (H), the number of subgraphs, the values for section ratios, and the number of trigrams, as demonstrated in Figure \ref{table12}. We keep the feature hasher size for the trigram feature constant, as its size is relatively small compared to other features. Among the experiments, case (2)—which incorporated two levels of in-flow and out-flow, ten subgraphs, a section ratio greater than 1.5, H with 100 features for subgraph opcode sequences, H with 100 features for subgraph API sequences, H with 100 features for each subgraph signature, H with 200 features for the whole CFG signature, and H with 20 features for fifteen trigrams—demonstrated superior performance across the evaluation metrics compared to the other cases, as indicated in Figure \ref{table12}. This answers the research question (R4).

\begin{figure}[!h]
    \centering
    \includegraphics[width=0.9\textwidth, trim = 0cm 0cm 0cm 0cm, clip]{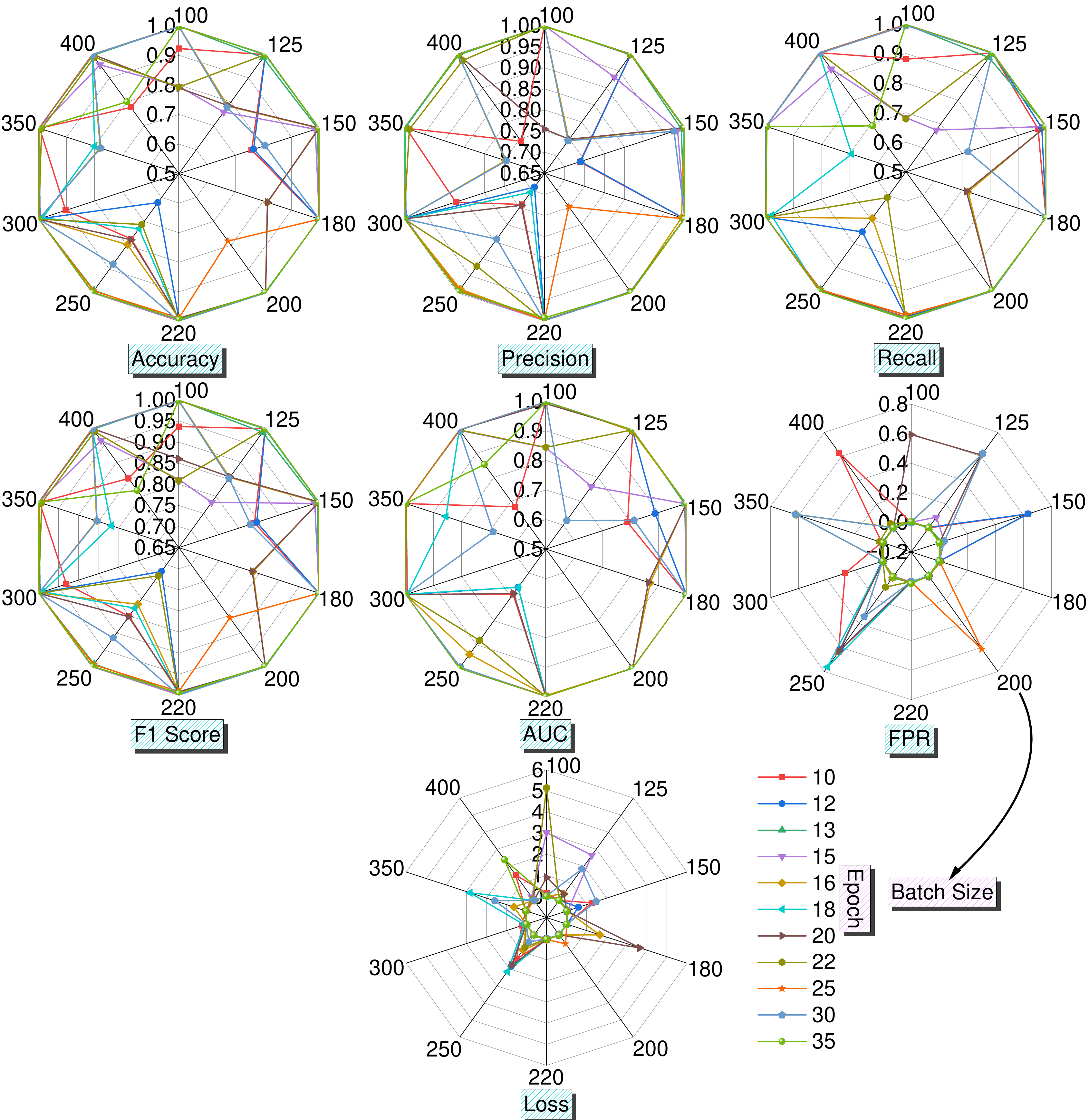}
    \caption{Experiments with Various Epochs and Batches.}
    \label{epochs_batches_fig}
\end{figure}
\textit{DNN Experiments with Different Epochs and Batches:} Epochs and batch sizes are crucial hyperparameters in training DNN as they significantly impact the model's performance and generalization capabilities. The epoch tells the number of times the model goes through the entire training dataset, while the batch size tells the number of samples the model processes before updating its weights. The optimal values of these two hyperparameters help prevent underfitting and overfitting, ensuring the model learns effectively while generalizing to unseen data. They also contribute to the stability of the training process and enhance computational efficiency. Therefore, we experiment by training the DNN13 model on all combinations of epoch and batch size from the epoch list = [10, 12, 13, 15, 16, 18, 20, 22, 25, 30, 35] and batch size list = [100, 125, 150, 180, 200, 220, 250, 300, 350, 400]. As shown in Figure \ref{epochs_batches_fig}, the combination of 12 epochs and a batch size of 200 yields the best performance across all combinations in terms of all evaluation metrics, including accuracy, precision, recall, AUC, F1-score, FPR, and loss.

\begin{figure}[t]
    \centering
    \includegraphics[width=0.7\textwidth, trim = 0cm 0cm 0cm 0cm, clip]{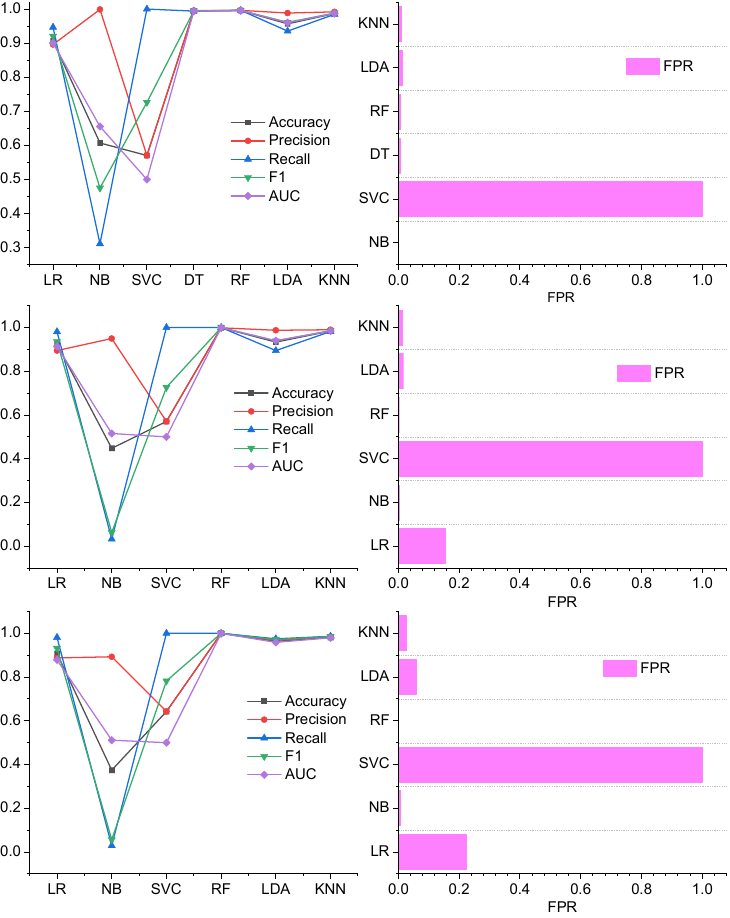}
    \caption{Experiments with Traditional ML Models.}
    \label{traditional_ML_Models}
\end{figure}

\textit{Experiments with Traditional ML Models:} We conduct experiments employing traditional ML models, including Logistic Regression (LR), Decision Trees (DT), Random Forest (RF), Support Vector Classifier (SVC), K-Nearest Neighbors (KNN), Naive Bayes (NB), and Linear Discriminant Analysis (LDA). These models are particularly effective with small datasets and offer key advantages such as providing insights into feature importance, fast training time, low complexity, and resource efficiency. Initially, we scale the advanced features before inputting them into these models. However, we observe that most models produce overly optimistic results across all evaluation metrics, raising concerns about overfitting and limited generalizability. To address this, we further evaluate the models using unscaled features. Despite this adjustment, the DT model continues to yield implausibly perfect results for features derived from preprocessing-2 and preprocessing-3, prompting its exclusion from the final comparison due to lack of reliability. As shown in Figure \ref{traditional_ML_Models}, the RF model demonstrates superior performance compared to other traditional ML models in terms of evaluation metrics, producing over 99\% accuracy, precision, recall, F1-score, AUC, and a mere 0.063\% FPR. These results exhibit negligible variation compared to the DNN13 model, further validating the effectiveness of the advanced features across both traditional ML models and neural networks.\\

\textit{Comparison with state-of-the-art models:} We compare the PotentRegion4MalDetect with state-of-the-art techniques, including all three analysis categories: static, dynamic, and hybrid analysis. The detailed comparisons in Table \ref{table8}, Table \ref{table9}, and Table \ref{table10} reveal the PotentRegion4MalDetect's superiority. Although some works, such as Abbas et al. \cite{ref11} and Deniz et al. \cite{refy22}, claim to have reasonable accuracy, they consider a smaller dataset for experimentation. In contrast, the PotentRegion4MalDetect, which leverages a larger and more diverse dataset, demonstrates exceptional performance in terms of accuracy and FPR. Notably, none of the SOTA models focus on the features of the potential malicious regions, a key strength for the PotentRegion4MalDetect.
\begin{table}[b]
\centering
\begin{threeparttable}
\caption{PotentRegion4MalDetect \textit{Vs.} Static Models}
\renewcommand{\tabcolsep}{1pt}
\begin{tabular}{|c|l|c|c|}
\hline
\textbf{Model}    & \multicolumn{1}{c|}{\textbf{Dataset}}                                                                                          & \textbf{Accuracy (\%)}                                                        & \textbf{FPR (\%)}                                                  \\ \hline
Sota et al. \cite{related1}          & \begin{tabular}[c]{@{}l@{}}8000B and 18000M\end{tabular} & \begin{tabular}[c]{@{}c@{}}95.1\end{tabular}                  & \begin{tabular}[c]{@{}c@{}}-\end{tabular} \\ \hline
Ömer et al. \cite{ref58}          & 45,306M                                                                                                                 & 97.78                                                                  & 2.98 \\ \hline
Sudan et al. \cite{ref59}          & \begin{tabular}[c]{@{}l@{}}5000B and 5000M\end{tabular}                 & 91.91                                                                  & -                                                             \\ \hline
Abbas et al. \cite{ref11}          & \begin{tabular}[c]{@{}l@{}}200B and 500M\end{tabular}                                                           & 98.25                                                                   &     -                                                \\ \hline
Deniz et al. \cite{refy22}          & \begin{tabular}[c]{@{}l@{}}986B and 1095M\end{tabular}                                                          & \begin{tabular}[c]{@{}c@{}}98.3\end{tabular} & -                                                \\ \hline
\textbf{PotentRegion4MalDetect} & \begin{tabular}[c]{@{}l@{}}\textbf{11089B and 9504M}\end{tabular}                                     & \textbf{99.95}                                                         & \textbf{0.064}                                                        \\ \hline
\end{tabular}
\label{table8}
\begin{tablenotes}
\item\begin{center}\scriptsize{B: Benign, M: Malware}\end{center}
\end{tablenotes}
\end{threeparttable}
\end{table}
\begin{table}[]
\centering
\begin{threeparttable}
\caption{PotentRegion4MalDetect \textit{Vs.} Dynamic Models}
\renewcommand{\tabcolsep}{1pt}
\begin{tabular}{|c|l|c|c|}
\hline
\textbf{Model}    & \multicolumn{1}{c|}{\textbf{Dataset}}                                             & \textbf{Accuracy (\%)}                                                                                 & \textbf{FPR (\%)} \\ \hline
Pengbin et al. \cite{related2} &  1686B and 24817M                                                             & \begin{tabular}[c]{@{}c@{}}96.38\end{tabular} &  -\\ \hline
Markus et al. \cite{related3} & \begin{tabular}[c]{@{}l@{}}5683B and 14679M\end{tabular}              & 91 &  13       \\ \hline
Shiva et al. \cite{refy221} & \begin{tabular}[c]{@{}l@{}}3282B and 4151M \end{tabular}              & 97.97 & -      \\ \hline
\textbf{PotentRegion4MalDetect} & \begin{tabular}[c]{@{}l@{}}\textbf{9504M and 11089B}\end{tabular} & \textbf{99.95}                                                                                  & \textbf{0.064}       \\ \hline
\end{tabular}
\label{table9}
\begin{tablenotes}
\item\begin{center}\scriptsize{B: Benign,M: Malware}\end{center}
\end{tablenotes}
\end{threeparttable}
\end{table}
\begin{table}[]
\centering
\begin{threeparttable}
\caption{PotentRegion4MalDetect \textit{Vs.} Hybrid Models}
\centering
\renewcommand{\tabcolsep}{2pt}
\begin{tabular}{|c|l|c|c|}
\hline
\textbf{Model}    & \multicolumn{1}{c|}{\textbf{Dataset}}                                                                   & \textbf{Accuracy (\%)} & \textbf{FPR (\%)}    \\ \hline
Asma et al. \cite{related5}           & \begin{tabular}[c]{@{}l@{}}4835B and 6877M\end{tabular} & 95            & 4               \\ \hline
Weijie et al. \cite{ref42}           & \begin{tabular}[c]{@{}l@{}}760 B and 3490 M \end{tabular} & 97.2            & -               \\ \hline
Asma et al. \cite{hyb1}           & \begin{tabular}[c]{@{}c@{}}\small{2000B and 3000M}\end{tabular} & 97.91 & 2.08 \\ \hline
\textbf{PotentRegion4MalDetect} & \begin{tabular}[c]{@{}l@{}}\textbf{11089B and 9504M} \end{tabular}               & \textbf{99.95}  & \textbf{0.064} \\ \hline
\end{tabular}
\label{table10}
\begin{tablenotes}
\item\begin{center}\scriptsize{B: Benign, M: Malware}\end{center}
\end{tablenotes}
\end{threeparttable}
\end{table}

\textit{Limitations of the Proposed Work:} The `\textit{PotentRegion4MalDetect}' model demonstrates impressive results compared to models focusing on the entire binary. However, a higher number of function calls in a binary may increase the computational time due to the usage of the DFS to find all possible paths from source to destination in mapping the APIs and malicious strings to the CFG nodes. Therefore, we limit our experimentation to binaries with less than 300 function calls to ensure a more efficient and manageable analysis. Nevertheless, future work may explore optimization strategies such as heuristic-based pruning and graph abstraction techniques to reduce the computational overhead associated with exhaustive path enumeration.
\section{Conclusion}
\label{section5}
Researchers focus on the entire binary for feature extraction instead of focusing on regions where maliciousness is present. This approach has a downside, i.e., attackers inject malicious code into a benign binary, making the model extract more benign features than malware to bypass the Machine Learning (ML) based approaches. However, we propose a novel approach to extract features from the most potentially malicious regions in a binary along with the features from the completely preprocessed Control Flow Graph (CFG). To see the robustness of the extracted advanced features, we compare these features with the features extracted from the whole binary. We observe a significant performance improvement by including advanced features. The advanced features produce 8.13\% and 1.44\% high SHapley Additive exPlanations (SHAP) Absolute Mean and Beeswarm values compared to the features extracted from the entire binary. We train and test the advanced features on a Deep Neural Network (DNN) with thirteen layers using various combinations of epochs and batch sizes. The combination of 12 epochs and a batch size of 200 yields the best results, achieving an accuracy of 99.95\%, precision of 99.97\%, recall of 99.97\%, AUC of 99.97\%, an F1 score of 99.98\%, and a False Positive Rate (FPR) of 0.063\%. Further, we check with traditional ML models such as LR, DT, RF, SVC, KNN, NB, and LDA and find that RF results exhibit negligible variation compared to the DNN13 model. Finally, we compare the proposed model with state-of-the-art static, dynamic, and hybrid models focused on the whole binary and show why focusing on potential malicious regions is a better approach.
\bibliography{casrefs}

\begin{thebibliography}{10}

\bibitem{later1}
Statista.
\newblock {Number of internet and social media users worldwide as of July
  2022}.
\newblock
  \url{https://www.statista.com/statistics/617136/digital-population-worldwide/}.
\newblock [Online; accessed 2022].

\bibitem{later3}
Clare Stouffer.
\newblock {115 cybersecurity statistics + trends to know in 2023}.
\newblock
  \url{https://us.norton.com/blog/emerging-threats/cybersecurity-statistics}.
\newblock [Online; accessed 2022].

\bibitem{iref1}
Statista.
\newblock {Percentage of organizations victimized by ransomware attacks
  worldwide from 2018 to 2022}.
\newblock
  \url{https://www.statista.com/statistics/204457/businesses-ransomware-attack-rate/}.
\newblock [Online; accessed 2022].

\bibitem{irefstat}
AV-ATLAS.
\newblock {Malware}.
\newblock \url{https://www.av-test.org/en/statistics/malware/}.
\newblock [Online; accessed 2024].

\bibitem{iref2}
AV-ATLAS.
\newblock {Statistics}.
\newblock \url{https://portal.av-atlas.org/malware/statistics}.
\newblock [Online; accessed 2022].

\bibitem{mld1}
Adeilson~Antonio da~Silva and Mauricio Pamplona~Segundo.
\newblock On deceiving malware classification with section injection.
\newblock {\em Machine Learning and Knowledge Extraction}, 5(1):144--168, 2023.

\bibitem{mld2}
Javier Yuste, Eduardo~G Pardo, and Juan Tapiador.
\newblock Optimization of code caves in malware binaries to evade machine
  learning detectors.
\newblock {\em Computers \& Security}, 116:102643, 2022.

\bibitem{related2}
Pengbin Feng, Le~Gai, Li~Yang, Qin Wang, Teng Li, Ning Xi, and Jianfeng Ma.
\newblock Dawngnn: Documentation augmented windows malware detection using
  graph neural network.
\newblock {\em Computers \& Security}, page 103788, 2024.

\bibitem{related3}
Ring Markus, Schl{\"o}r Daniel, Wunderlich Sarah, Landes Dieter, and Hotho
  Andreas.
\newblock Malware detection on windows audit logs using lstms [j].
\newblock {\em Computers \& Security, 2021 (prepublish)}, 2021.

\bibitem{related4}
Pascal Maniriho, Abdun~Naser Mahmood, and Mohammad Jabed~Morshed Chowdhury.
\newblock Memaldet: A memory analysis-based malware detection framework using
  deep autoencoders and stacked ensemble under temporal evaluations.
\newblock {\em Computers \& Security}, 142:103864, 2024.

\bibitem{related5}
Asma~A Alhashmi, Abdulbasit~A Darem, Sultan~M Alanazi, Abdullah~M Alashjaee,
  Bader Aldughayfiq, Fuad~A Ghaleb, Shouki~A Ebad, and Majed~A Alanazi.
\newblock Hybrid malware variant detection model with extreme gradient boosting
  and artificial neural network classifiers.
\newblock {\em Computers, Materials \& Continua}, 76(3), 2023.

\bibitem{ref42}
Weijie Han, Jingfeng Xue, Yong Wang, Zhenyan Liu, and Zixiao Kong.
\newblock Malinsight: A systematic profiling based malware detection framework.
\newblock {\em Journal of Network and Computer Applications}, 125:236--250,
  2019.

\bibitem{ref23}
Seungho Jeon and Jongsub Moon.
\newblock Malware-detection method with a convolutional recurrent neural
  network using opcode sequences.
\newblock {\em Information Sciences}, 535:1--15, 2020.

\bibitem{ref11}
Abbas Yazdinejad, Hamed HaddadPajouh, Ali Dehghantanha, Reza~M Parizi, Gautam
  Srivastava, and Mu-Yen Chen.
\newblock Cryptocurrency malware hunting: A deep recurrent neural network
  approach.
\newblock {\em Applied Soft Computing}, 96:106630, 2020.

\bibitem{Sibelg}
Sibel G{\"u}lmez and Ibrahim Sogukpinar.
\newblock Graph-based malware detection using opcode sequences.
\newblock In {\em 2021 9th International Symposium on Digital Forensics and
  Security (ISDFS)}, pages 1--5. IEEE, 2021.

\bibitem{ref010}
Sudan Jha, Deepak Prashar, Hoang~Viet Long, and David Taniar.
\newblock Recurrent neural network for detecting malware.
\newblock {\em computers \& security}, 99:102037, 2020.

\bibitem{related1}
Sota Okubo, Tomotaka Kimura, and Jun Cheng.
\newblock Entropy-based malware detection using one dimensional cnn.
\newblock In {\em 2024 International Conference on Consumer Electronics-Taiwan
  (ICCE-Taiwan)}, pages 763--764. IEEE, 2024.

\bibitem{refy22}
Den{\i}z Dem{\i}rc{\i}, Cengiz Acarturk, et~al.
\newblock Static malware detection using stacked bilstm and gpt-2.
\newblock {\em IEEE Access}, 10:58488--58502, 2022.

\bibitem{Yifeij}
Yifei Jian, Hongbo Kuang, Chenglong Ren, Zicheng Ma, and Haizhou Wang.
\newblock A novel framework for image-based malware detection with a deep
  neural network.
\newblock {\em Computers \& Security}, 109:102400, 2021.

\bibitem{Kohei}
Kohei Tsunewaki, Tomotaka Kimura, and Jun Cheng.
\newblock Two-stage malware detection using calling api-based bilstm.
\newblock In {\em 2023 International Conference on Consumer Electronics-Taiwan
  (ICCE-Taiwan)}, pages 471--472. IEEE, 2023.

\bibitem{refsyscall}
Smita Naval, Vijay Laxmi, Muttukrishnan Rajarajan, Manoj~Singh Gaur, and Mauro
  Conti.
\newblock Employing program semantics for malware detection.
\newblock {\em IEEE Transactions on Information Forensics and Security},
  10(12):2591--2604, 2015.

\bibitem{Hongbi}
Hongbi Kim and Taejin Lee.
\newblock Research on autoencdoer technology for malware feature purification.
\newblock In {\em 2021 21st ACIS International Winter Conference on Software
  Engineering, Artificial Intelligence, Networking and Parallel/Distributed
  Computing (SNPD-Winter)}, pages 236--239. IEEE, 2021.

\bibitem{Vidhi}
Vidhi Garg and Rajesh~Kumar Yadav.
\newblock Malware detection based on api calls frequency.
\newblock In {\em 2019 4th International Conference on Information Systems and
  Computer Networks (ISCON)}, pages 400--404. IEEE, 2019.

\bibitem{ref10}
Mandiant.
\newblock {StringSifter}.
\newblock \url{https://github.com/mandiant/stringsifter}.
\newblock [Online; accessed 2021].

\bibitem{ref78}
Daniel Bilar.
\newblock Opcodes as predictor for malware.
\newblock {\em International journal of electronic security and digital
  forensics}, 1(2):156--168, 2007.

\bibitem{ref13}
Hyrum~S Anderson and Phil Roth.
\newblock Ember: an open dataset for training static pe malware machine
  learning models.
\newblock {\em arXiv preprint arXiv:1804.04637}, 2018.

\bibitem{ref30}
Ben Athiwaratkun and Jack~W Stokes.
\newblock Malware classification with lstm and gru language models and a
  character-level cnn.
\newblock In {\em 2017 IEEE international conference on acoustics, speech and
  signal processing (ICASSP)}, pages 2482--2486. IEEE, 2017.

\bibitem{ref29}
Moustafa Saleh, E~Paul Ratazzi, and Shouhuai Xu.
\newblock A control flow graph-based signature for packer identification.
\newblock In {\em MILCOM 2017-2017 IEEE Military Communications Conference
  (MILCOM)}, pages 683--688. IEEE, 2017.

\bibitem{refpractice}
Michael Sikorski and Andrew Honig.
\newblock {\em Practical malware analysis: the hands-on guide to dissecting
  malicious software}.
\newblock no starch press, 2012.

\bibitem{msg_rama}
Rama~Krishna Koppanati and Sateesh~K Peddoju.
\newblock Msg: Missing-sequence generator for metamorphic malware detection.
\newblock {\em Journal of Information Security and Applications}, 89:103962,
  2025.

\bibitem{overfast}
Christian Garbin, Xingquan Zhu, and Oge Marques.
\newblock Dropout vs. batch normalization: an empirical study of their impact
  to deep learning.
\newblock {\em Multimedia Tools and Applications}, 79:12777--12815, 2020.

\bibitem{featurehasher}
{Feature Hasher}.
\newblock
  \url{https://scikit-learn.org/stable/modules/generated/sklearn.feature_extraction.FeatureHasher.html}.
\newblock [Online; accessed 2020].

\bibitem{ref7}
VirusTotal.
\newblock {}.
\newblock \url{https://www.virustotal.com/}.
\newblock [Online; accessed 2021].

\bibitem{refdata}
bormaa.
\newblock {Benign-NET files}.
\newblock \url{https://github.com/bormaa/Benign-NET}.
\newblock [Online; accessed 2022].

\bibitem{ref58}
{\"O}mer Aslan and Abdullah~Asim Yilmaz.
\newblock A new malware classification framework based on deep learning
  algorithms.
\newblock {\em Ieee Access}, 9:87936--87951, 2021.

\bibitem{ref59}
Sudan Jha, Deepak Prashar, Hoang~Viet Long, and David Taniar.
\newblock Recurrent neural network for detecting malware.
\newblock {\em computers \& security}, 99:102037, 2020.

\bibitem{refy221}
Shiva~Darshan SL and CD~Jaidhar.
\newblock Windows malware detector using convolutional neural network based on
  visualization images.
\newblock {\em IEEE Transactions on Emerging Topics in Computing},
  9(2):1057--1069, 2019.

\bibitem{hyb1}
Asma~A Alhashmi, Abdulbasit~A Darem, Abdullah~M Alashjaee, Sultan~M Alanazi,
  Tareq~M Alkhaldi, Shouki~A Ebad, Fuad~A Ghaleb, and Aloyoun~M Almadani.
\newblock Similarity-based hybrid malware detection model using api calls.
\newblock {\em Mathematics}, 11(13):2944, 2023.

\end{thebibliography}

\end{document}